\documentclass[]{pasj01}

\Received{}
\Accepted{}
 
\usepackage{hyphenat} 
\usepackage{url}
\usepackage{comment}
    \makeatletter
\def\@fnsymbol#1{\ensuremath{\ifcase#1\or \dagger\or \ddagger\or
   \mathsection\or \mathparagraph\or \|\or **\or \dagger\dagger
   \or \ddagger\ddagger \else\@ctrerr\fi}}
    \makeatother
 
\begin{document} 

\title{ 
Subaru Hyper Suprime-Cam revisits the large-scale environmental dependence on galaxy morphology over 360 deg$^2$ at z=0.3--0.6}


\author{Rhythm \textsc{Shimakawa}\altaffilmark{1,2}$^\ast$\thanks{NAOJ Fellow}}
\altaffiltext{1}{National Astronomical Observatory of Japan (NAOJ), National Institutes of Natural Sciences, Osawa, Mitaka, Tokyo 181-8588, Japan}
\email{rhythm.shimakawa@nao.ac.jp (RS)}

\author{Takumi S. \textsc{Tanaka}\altaffilmark{1,2}}
\altaffiltext{2}{Department of Astronomy, School of Science, The University of Tokyo, 7-3-1 Hongo, Bunkyo-ku, Tokyo 113-0033, Japan}

\author{Seiji \textsc{Toshikage}\altaffilmark{3}}
\altaffiltext{3}{Astronomical Institute, Tohoku University, 6-3, Aramaki, Aoba, Sendai, Miyagi, 980-8578, Japan}

\author{Masayuki \textsc{Tanaka}\altaffilmark{1}}



\maketitle

\begin{abstract}
This study investigates the role of large-scale environments on the fraction of 
spiral galaxies at $z=$ 0.3--0.6 sliced to three redshift bins of $\Delta z=0.1$. 
Here, we sample 276~220 massive galaxies in a limited stellar mass of 
$5\times10^{10}$ solar mass ($\sim M^\ast$) over 360 deg$^2$, as obtained from the 
Second Public Data Release of the Hyper Suprime-Cam Subaru Strategic Program 
(HSC-SSP). 
By combining projected two-dimensional density information (Shimakawa et al. 2021) 
and the CAMIRA cluster catalog (Oguri et al. 2018), we investigate the spiral 
fraction across large-scale overdensities and in the vicinity of red sequence 
clusters. 
We adopt transfer learning to significantly reduce the cost of labeling spiral 
galaxies and then perform stacking analysis across the entire field to overcome the
limitations of sample size. Here we employ a morphological classification catalog 
by the Galaxy Zoo Hubble \citep{Willett2017} to train the deep learning model. 
Based on 74~103 sources classified as spirals, we find moderate 
morphology--density relations on ten comoving Mpc scale, thanks to the wide-field 
coverage of HSC-SSP. Clear deficits of spiral galaxies have also been confirmed, 
in and around 1136 red sequence clusters. 
Furthermore, we verify whether there is a large-scale environmental dependence on 
rest-frame $u-r$ colors of spiral galaxies; however, such a tendency was not 
observed in our sample. 
\end{abstract}


\section{Introduction}

Physical mechanisms of environmental impacts on galaxy formation and evolution, 
and even the very existence of such effects, have been long-standing controversial 
topics, depending on such an evolutionary phase, a cosmic time, or an 
environmental scale (e.g., reviews by \cite{Boselli2006,Blanton2009}; theoretical 
work by 
\cite{Davis1985,Dekel1986,Moore1998,White1991,Keres2005,Bekki2011,Somerville2015,Joshi2020,Donnari2021}). 
For example, many studies have reported a close color--magnitude relationship, 
called red sequence, since redshift $\sim2$ or less 
\citep{Vaucouleurs1961,Visvanathan1977,Butcher1984,Bower1992,Bower1998,Terlevich2001,Gladders2005,White2005,Kodama2007,Cooper2007,Muzzin2008,Mei2009,Rykoff2014,Oguri2018}. 
Such color dependencies on the local galaxy over-density have been verified in more 
diversified environments via extensive wide-field surveys covering more than a 
hundred square degrees 
\citep{Balogh2004,Tanaka2005,Baldry2006,Skibba2009,Peng2010,Alpaslan2015,Costa-Duarte2018}.
Indeed, less active star formations of cluster galaxies have been detected in 
various observations, which suggests that dense environments induce star formation 
quenching 
\citep{Kauffmann2004,Poggianti2006,Wetzel2012,Wetzel2013,Muzzin2014,Old2020}. 
Environmental quenching is often regarded as a quenching process, together with 
mass-dependent star formation quenching \citep{Peng2010,Grutzbauch2011,Peng2012}. 

The morphology--density relation, which is the primary focus of this paper, 
is a more contentious research topic than the environmental dependence of galaxy 
colors, which refers to the increase in disk early-type (S0) galaxies in dense 
regions, at the cost of a fraction of spiral galaxies 
\citep{Dressler1980,Larson1980,Dressler1997,Couch1998,Goto2003a,Smith2005,Postman2005,Capak2007,Guzzo2007,Wel2008,Poggianti2009,Tasca2009,Cappellari2011,Lietzen2012,Houghton2013,Fogarty2014}. 
The relationship apparently indicates a morphological transition of star-forming 
galaxies to the early-type galaxies owing to environmental effects like ram 
pressure stripping and gas strangulation 
\citep{Moran2007,Wolf2009,Kovac2010,Vulcani2011}. 
However, previous studies have reported that such an apparent environmental 
dependence tends to almost disappear at a fixed stellar mass, especially at the 
high mass end, thus suggesting that the mass would primarily control such an 
observational trend (e.g. \cite{Holden2007,Bamford2009}, and see also 
\cite{Brough2017,Veale2017,Greene2017} for the kinematic morphology--density 
relation in early-type galaxies). 

Moreover, quite importantly, a significant fluctuation in the environmental 
dependence would exist, just as there is a great diversity of galaxy clusters 
\citep{Rood1971,Butcher1984,Goto2005,Hashimoto2019}. Recent wide-field deep legacy 
surveys have determined more than a thousand cluster samples by systematic cluster 
search out to $z\sim1$, e.g., CAMIRA \citep{Oguri2018}, AMICO \citep{Maturi2019}, 
and X-CLASS \citep{Koulouridis2021}. 
Besides, we have free access to the wide-field projected density map at $z<1$ 
over 360 deg$^2$ \citep{Shimakawa2021}. These libraries that sufficiently 
cover the sample size and survey volume will enable us to obtain a more systematic 
perspective on environmental dependence across a formation history of the 
large-scale structure. 

Therefore, this motivates us to revisit the morphology--density relation using 
large data-sets based on wide-field deep data. Specifically, we obtain the sample 
from the Second Public Data Release of the Hyper Suprime-Cam Subaru Strategic 
Program (\cite{Aihara2019}; hereafter referred to as the HSC-SSP PDR2), as well as 
the density map \citep{Shimakawa2021} and CAMIRA cluster catalog \citep{Oguri2018}, 
both of which are based on the HSC-SSP PDR2 library (\S\ref{s2}). Furthermore, this 
research employs a novel approach, transfer learning 
\citep{Bozinovski1976,Pratt1993}, to achieve an effective search of spiral galaxies 
from big data (\S\ref{s3}). 
Transfer learning is a machine-learning technique occasionally used for the 
classifications of galaxy mergers \citep{Ackermann2018} and radio galaxies 
\citep{Tang2019}, including the application of a visual morphology classification 
to another survey \citep{Dominguez2019}. Based on the deep learning classification, 
we investigate the environmental dependence of spiral fractions on 10 comoving Mpc 
(co-Mpc) scale and in the vicinity of galaxy clusters (\S\ref{s4}). This study 
primarily focuses on such a large-scale environmental effect by taking advantage of 
the wide-field coverage (360 deg$^2$) and owing to photometric redshift errors on 
the backside. Finally, we discuss obtained results and summarize this research in 
the last section (\S\ref{s5} and \S\ref{s6}). 

Similar to our previous work \citep{Shimakawa2021}, this study adopts the AB 
magnitude system \citep{Oke1983} and the \citet{Chabrier2003} initial mass function. 
We also assume cosmological parameters, $\Omega_M=0.279$, $\Omega_\Lambda=0.721$, 
and $h=0.7$, in a flat lambda cold dark matter model, which are consistent with 
those from the WMAP nine-year data \citep{Hinshaw2013}. 
When referring to our figures or tables, we designate their initials by capital 
letters (e.g., Fig.~1, Table~1, Section~1) to avoid confusion with those in the 
literature (e.g., fig.~1, table~1, or section~1).

\section{Data}\label{s2}

This section overviews an original sample obtained from the HSC-SSP PDR2, which is 
applied in the density measurement in our previous work \citep{Shimakawa2021}. We 
then examine how to prepare an imaging data-set for selecting spiral galaxies from 
our sample. 

\subsection{HSC-SSP PDR2}\label{s21}

\begin{figure}
\begin{center}
	\includegraphics[width=7.5cm]{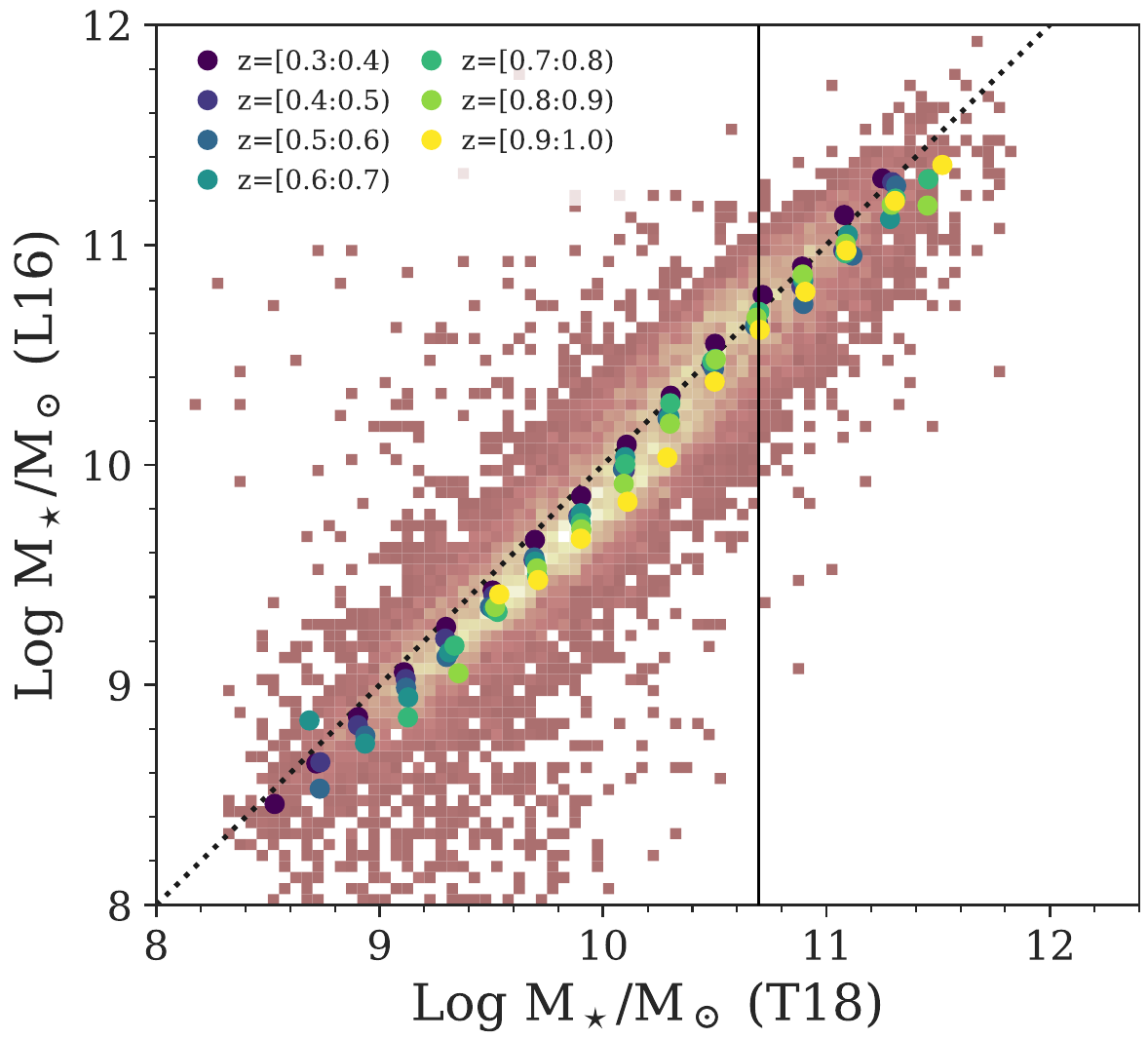}
\end{center}
    \caption{
    Comparisons of stellar mass estimates of $i<23$ galaxies at $z=[0.3:1.0)$ by 
    (x-axis) \citet[based on HSC $grizy$ bands]{Tanaka2018} with those by 
    (y-axis) \citet{Laigle2016}, which use all photometric bands available in 
    the COSMOS field \citep{Scoville2007}. 
    The median values are depicted by the circles (colors of the symbols vary 
    depending on the redshift bin as denoted in the legend).
    The black solid vertical line indicates the stellar mass limit 
    (M$_\star=5\times10^{10}$ M$_\odot$) determined in this work. 
    }
    \label{fig1}
\end{figure}

This work is based on data, all of which have been available in the public 
database of the HSC-SSP PDR2\footnote{\url{https://hsc-release.mtk.nao.ac.jp}} 
since May 30, 2019 \citep{Aihara2019}. The database includes all data taken from 
March 2014 to January 2018 (refer to \cite[table 2]{Aihara2019} for further 
details) with the Hyper Suprime-Cam on the 8.2 m Subaru Telescope 
\citep{Miyazaki2018,Furusawa2018,Kawanomoto2018,Komiyama2018}. 
Using the database, we can access the science-ready source catalog and 
reduced fits data generated by the dedicated pipeline termed {\tt hscPipe} 
\citep{Bosch2018}. 

Throughout this study, we employ only luminous galaxies in a limited stellar mass 
of M$_\star=5\times10^{10}$ solar mass at a photometric redshift range of 
$z=[0.3:0.6)$. 
The sample is originally based on approximately eight million $i$-band magnitude 
limited ($<23$ mag) sources of the study by \citet{Shimakawa2021}. They are 
detected in all HSC bands ($grizy$) with more than five sigma detections in the 
{\tt cmodel} measurement (see 
\cite[\S4.9.9]{Bosch2018} and \cite[\S3.1]{Abazajian2004} for details of the 
{\tt cmodel} magnitude). Suspicious detections are excluded using the following 
flags ({\tt is False}) on the HSC-SSP PDR2 database: 
{\tt pixelflags\_edge}, {\tt pixelflags\_interpolatedcenter}, 
{\tt pixelflags\_saturatedcenter}, {\tt pixelflags\_crcenter}, 
{\tt pixelflags\_bad}, {\tt pixelflags\_bright\_objectcenter},  
{\tt mask\_s18a\_bright\_objectcenter}, and {\tt i\_sdsscentroid\_flag}
(\cite{Coupon2018,Aihara2019}; \cite[table 2]{Bosch2018}). 

Photometric redshifts and stellar masses of the sample are derived from 
{\tt Mizuki}, a SED-based photo-$z$ code \citep{Tanaka2015}. The biweight 
dispersion of $\Delta z=|z_\mathrm{spec}-z_\mathrm{photo}|$ is $\leq0.04$, and 
the outlier rate ($|\Delta z|>0.15$) is $\sim0.1$ for the HSC-SSP sources at 
$i<23$ mag (see \cite[table 2]{Tanaka2018} for more details). 
We solely utilize samples that satisfy the reduced chi-square $\chi_\nu<5$ of the 
best-fitting model, following the recommendation in \citet[section 7]{Tanaka2018}. 
For the sample at redshift $z=[0.4:0.5)$, we set a further selection threshold 
to minimize an uncertainty from a Balmer--Lyman break degeneracy. Since 
misclassified samples at $z\sim3$ tend to exhibit very young ages ($\sim1$ Gyr) 
with high specific star-formation rates ($>1$ Gyr$^{-1}$) 
in the {\tt Mizuki}, we discard such sources. 

In addition, we conduct a sanity check on the stellar mass measurement by 
{\tt Mizuki} at $z=[0.3:1.0)$ using only the $grizy$ photometry available in 
HSC-SSP as compared to those by COSMOS2015 \citep{Laigle2016}, which are based on 
34-band photometry (Fig.~\ref{fig1}). 
The HSC-only measurement restricted to the optical regime generally tends to 
overestimate stellar masses of galaxies at higher redshifts ($z\gtrsim0.6$) and 
lower stellar masses ($\lesssim$ a few $10^{10}$ solar mass). However, 
Fig.~\ref{fig1} shows that we can obtain logical values for massive galaxies 
at $z\lesssim0.6$, whose rest-frame optical wavelength can be covered optimally by 
$zy$ filters. We therefore decide to adopt massive galaxies at $z=[0.3:0.6)$, with 
stellar masses greater than $5\times10^{10}$ solar mass, which approximately 
corresponds to the characteristic stellar mass at that redshift 
\citep{Ilbert2010,Davidzon2017}. 
There are 276~220 galaxies that meet these requirements. Such thresholds are also 
selected from the seeing limit perspective (FWHM $=0.7$ arcsec). 
It is substantially challenging for the seeing-limited data to resolve the 
morphology of more compact (i.e., lower-mass) galaxies and higher-$z$ sources.

\subsection{Density map and galaxy cluster sample}\label{s22}

\begin{figure}
\begin{center}
	\includegraphics[width=8cm]{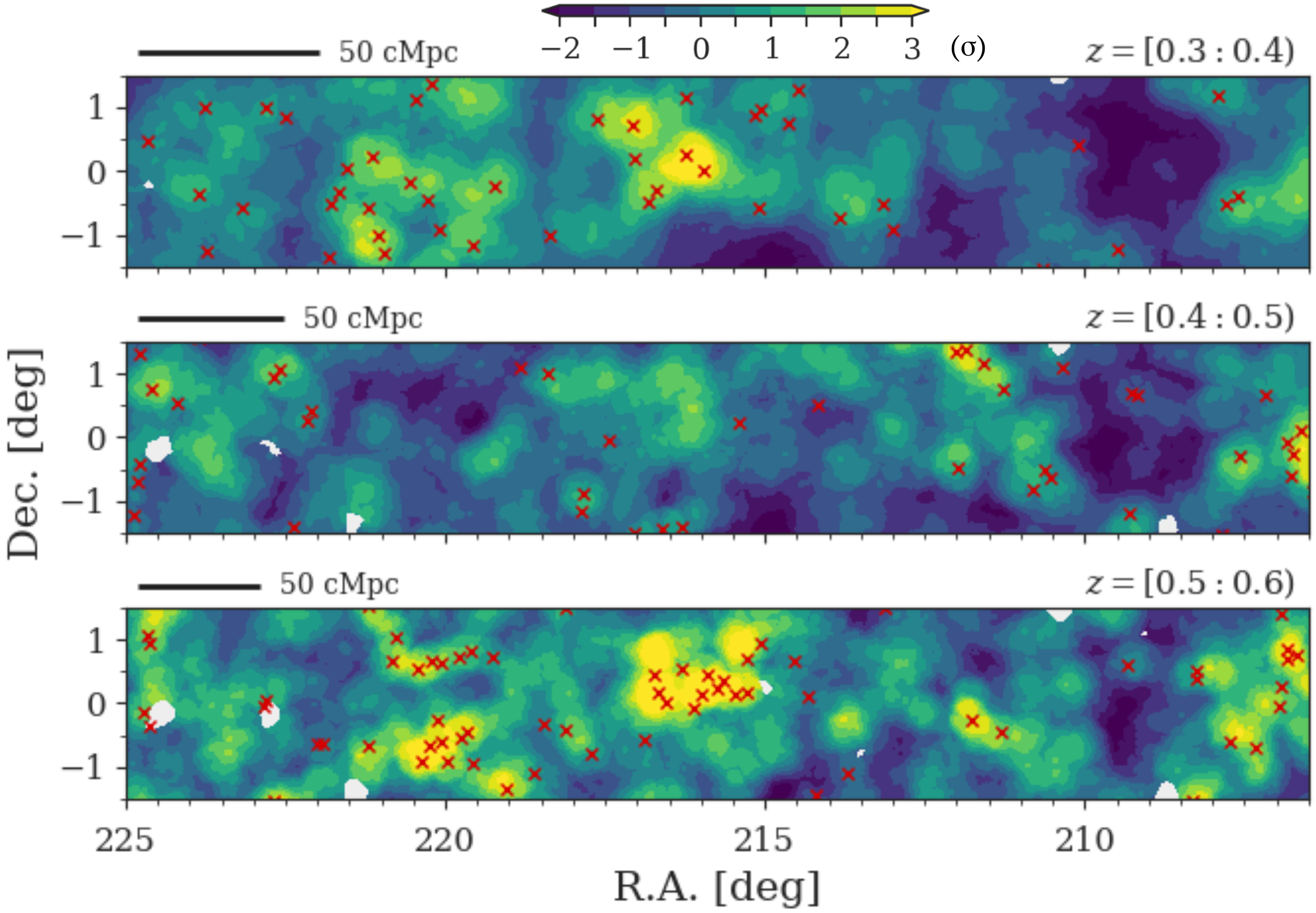}
\end{center}
    \caption{
    Examples of the density map (see \cite{{Shimakawa2021}} for the full version). 
    From top to bottom, projected two-dimensional overdensities at $z=$ [0.3:0.4), 
    [0.4:0.5), and [0.6:0.7), respectively. The color map depicts the density 
    excess in standard deviation ($\sigma_\mathrm{r=10Mpc}$). The overdensity is 
    measured for $i$-band magnitude limited galaxies ($i<23$) based on $r=10$ 
    co-Mpc aperture. 
    The red cross symbols indicate positions of the CAMIRA clusters at each 
    redshift slice \citep{Oguri2018}.
    While the CAMIRA clusters generally follow the large-scale overdensities, 
    there are some mismatch because measuring scales and tracers are different in 
    the two density measurements (see \cite{Shimakawa2021} for more details).
    }
    \label{fig2}
\end{figure}

We select an area of 360 deg$^2$ from the HSC-SSP Wide layer 
\citep[see figure 1]{Aihara2019}, where the projected 2D density maps are available 
\citep{Shimakawa2021}. In practice, the survey fields are organized by five large 
areas, one of which is illustrated in Fig.~\ref{fig2} as an example. 
\citet{Shimakawa2021} selected the survey area in an $i$-band limiting 
magnitude ($5\sigma$) of point-spread-function (PSF) in two arcsec diameters deeper 
than 26 mag. They also discard problematic fields adversely affected by bright 
stars ($<17.5$ mag) at a rate of more than $50$ percent (see section~2 in the 
literature). 
We verify that our mask correction provides the number density required at the 
target redshift range $z=[0.3:0.6)$, which is consistent with that based on the 
COSMOS2015 catalog \citep{Laigle2016}. Detailed information on the density 
estimation can be found in \citet[section 3.1]{Shimakawa2021}. To derive the 
overdensity of our sample, we consider values of the density excess using $r=10$ 
co-Mpc apertures ($\sigma_\mathrm{r=10Mpc}$) in three redshift slices of 
$z=[0.3:0.4)$, $[0.4:0.5)$, and $[0.5:0.6)$. Both the galaxy sample and density 
map are originally based on a study by \citet{Shimakawa2021}. Hence, the survey 
areas are completely consistent with each other. 

In addition, we employ a red sequence cluster catalog obtained from 
\citet{Oguri2018} to investigate spiral fractions of cluster neighborhoods. The 
cluster catalog was constructed based on the CAMIRA (Cluster-finding Algorithm 
based on Multi-band Identification of Red-sequence gAlaxies) algorithm 
\citep{Oguri2014}. 
The program estimates the 3D richness of red sequence galaxies in a limiting 
stellar mass of $10^{10.2}$ M$_\odot$ within one $h^{-1}$ physical Mpc (see 
\cite{Oguri2014} for more detailed information). We use their catalog correlating 
to the HSC-SSP PDR2, which includes the incremental update from the original 
version in the literature. We here adopt only cluster samples above the richness 
limit of 15, which roughly corresponds to the cluster virial mass of 
$\gtrsim10^{14}~h^{-1}$M$_\odot$ \citep{Oguri2018}. 
Such red sequence cluster samples will enable us to revisit the well-known 
morphology--density relation \citep{Dressler1980} in and around galaxy clusters 
at intermediate redshifts in a more inclusive manner.

\subsection{Image preparation}\label{s23}

This research utilizes gray-scale images from $i$-band data, consisting of 
relatively homogeneous and better-seeing data (FWHM $\sim0.55$ arcsec) than those 
in the other photometric bands (see e.g., \cite[fig.~2]{Shimakawa2021}). The HSC 
$i$-band filter captures the rest-frame 4400--6500 \AA\ of our targets at 
$z=[0.3:0.6)$. 
Beyond the technical reason above, we prefer to adopt the gray-scale than the 
multi-band colored image to avoid a color--morphology dependency (e.g., 
\cite{Vaucouleurs1961,Schawinski2014}) that may produce a classification bias. 
We employ samples with seeing FWHM $<0.7$ arcsec and then 
perform Gaussian smoothing for every data to match their seeing sizes to FWHM 
$=0.7$ arcsec, based on the seeing information in the corresponding {\tt patch} of 
the HSC-SSP PDR2 library. 

We convert $i$-band data of the targets into the machine-friendly gray-scale 
images by following the form of the arcsinh stretch, 
$F(x)\equiv\mathrm{arcsinh}(x/\beta)$, proposed by \citet{Lupton1999,Lupton2004}, 
where $x$ is a flux count per pixel in the $i$-band, and $\beta$ is the softening 
parameter (in this work, $\beta=1\times10^{-3}$). The zero-point magnitude of the 
$i$-band data is scaled to 19 mag (default in 
hscMap\footnote{\url{https://hsc-release.mtk.nao.ac.jp/hscMap-pdr2/app/}}) 
from the original value in HSC-SSP (27 mag). 
A normalization formula is described as follows, 
\begin{equation}
    f(x) = \left\{ \begin{array}{ll}
    0 & (x<m) \\
    F(x-m)/F(M-m) & (m\leq x\leq M) \\
    1 & (M<x),
  \end{array} \right.
\label{eq1}
\end{equation}
where $m$ is an undercut value ($m=5\times10^{-5}$) and $M$ represents an upper 
limit defined to be the maximum count within the central region of individuals 
($3\times3$ pixels) in this study. Subsequently, we cut out and convert 
the normalized data into the {\tt png} format with the image size of ($64\times64$) 
pixels (corresponding to $10.75\times10.75$ arcsec$^2$), to turn the data cube 
into a deep learning model (\S\ref{s31}). Examples of normalized images can be 
found in Fig.~\ref{fig6} and \ref{fig7} in the next section.

\section{Galaxy classification}\label{s3}

In this section, we first introduce our machine-learning-based approach 
(\S\ref{s31}) and explain how to train and validate the classification model by 
referencing it to the archive catalog, Galaxy Zoo Hubble \citep{Willett2017} in 
\S\ref{s32}. We then select spiral galaxies by applying the trained model to the 
normalized cutouts of the main targets (\S\ref{s33}). 

\subsection{Transfer learning}\label{s31}

\begin{figure}
\begin{center}
	\includegraphics[width=8cm]{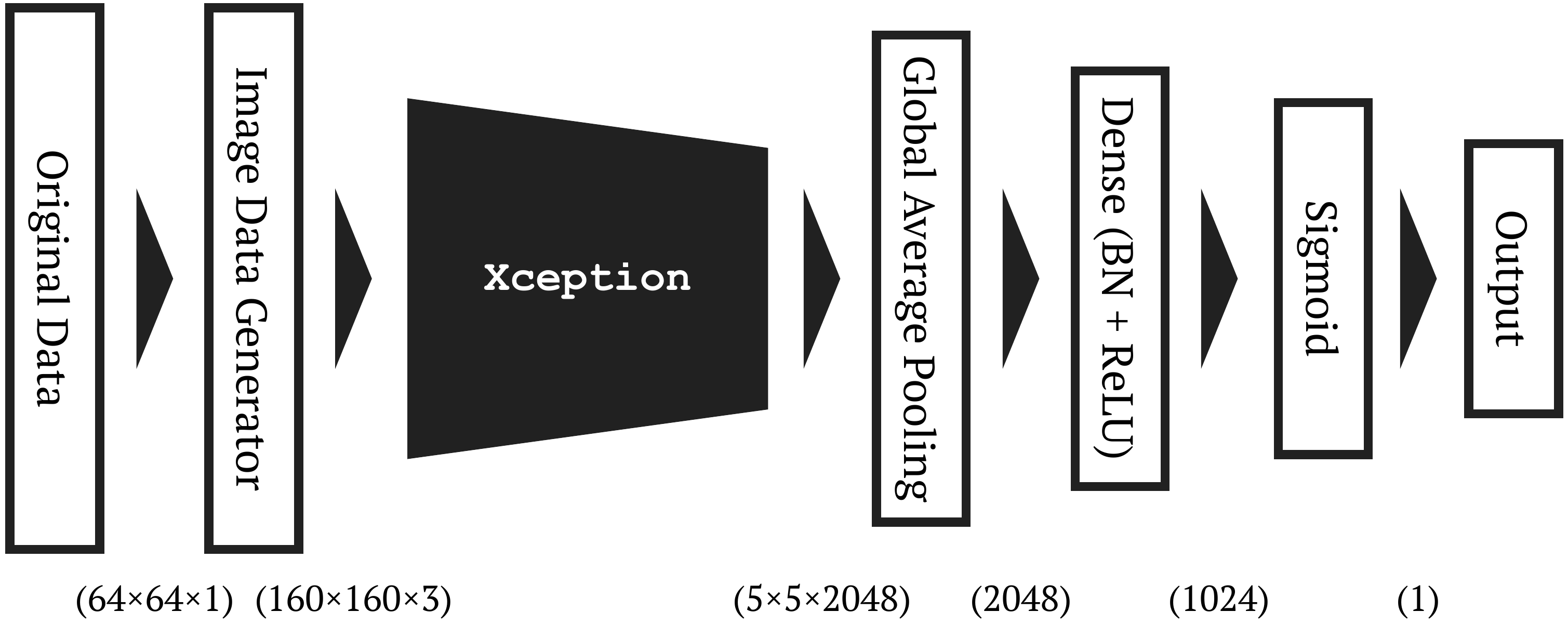}
\end{center}
    \caption{
    Image illustration of our transfer learning model architecture where a 
    two-class sigmoid classifier is attached to the {\tt Xception} model with 36 
    convolutional layers \citep{Chollet2016}. The parenthesized numbers indicate 
    the size of the data cube in each process. ``BN" means a batch normalization 
    layer \citep{Ioffe2015}. 
    }
    \label{fig3}
\end{figure}

We applied transfer learning (see, e.g., \cite{Bozinovski1976,Pratt1993} and 
a review by \cite{Tan2018}) to select spiral galaxies from the entire sample. 
Transfer learning is a machine learning technique, where a model pre-trained on 
one domain is diverted to another task. Modern deep learning models preliminarily 
trained with large imaging data cubes can be considered experts of feature 
recognition, delivering superior performance on galaxy images by including an 
adapted classifier, even if they are total strangers to the models. The major 
advantage of adopting transfer learning is that we can significantly reduce the 
sample size of the training data-set while keeping a robust classification 
accuracy. This approach has recently garnered increasing applications in the field 
of galaxy morphology classifications for various scientific purposes 
\citep{Ackermann2018,Tang2019,Dominguez2019}. 

In practice, we attached a two-class sigmoid classifier (0: non-spirals vs. 1: 
spirals) to the {\tt Xception} model \citep{Chollet2016} trained with the ImageNet 
data-set beforehand \citep{Russakovsky2014}. {\tt Xception} is a modern 
architecture of a convolutional neural network (CNN; \cite{Lecun1998,Lecun2015}) 
developed from the Inception architecture known as GoogLeNet \citep{Simonyan2014}. 
Throughout this study, we applied the deep learning model using {\tt TensorFlow} 
(version 2.2.0; \cite{Abadi2016}) and {\tt Keras} (version 2.3.0; 
\cite{Chollet2015}), under a single GPU, NVIDIA TITAN RTX. When we seeded random 
number generators built into programs, we adopted a fixed seed number of $30$. 
While different initial setups such as random seeds and GPUs would trigger a 
negligible mismatch ($\lesssim10$ percent, as far as we can ascertain) of spiral 
galaxy classification among models, the statistical trends discussed in this 
paper will not change (see, e.g., \cite{Pietrowski2021} about hardware-derived 
uncertainties). 

\begin{figure}
\begin{center}
	\includegraphics[width=7.5cm]{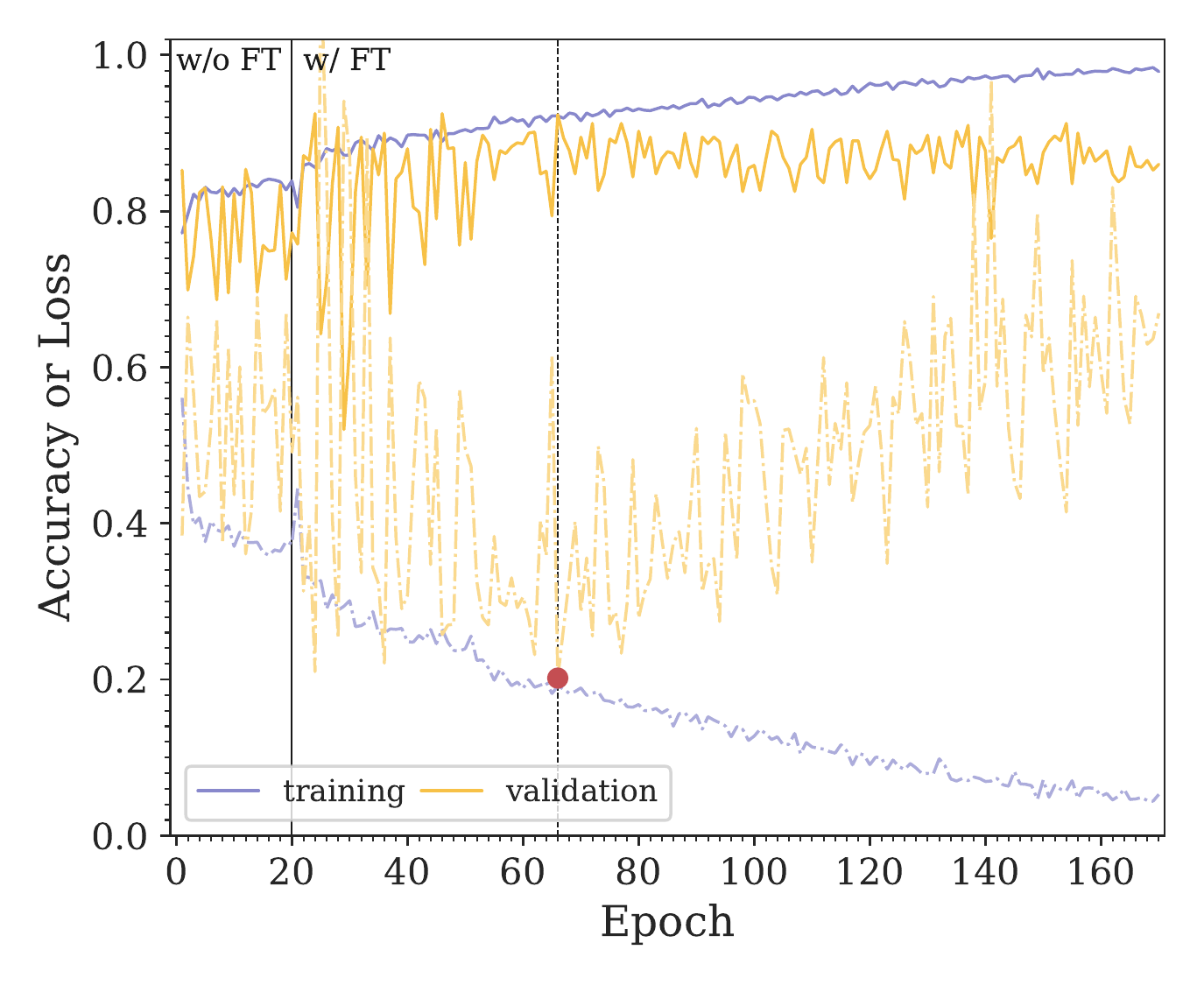}
\end{center}
    \caption{
    Training histories of the model accuracy (Eq.~\ref{eq2}; solid lines) and 
    binary cross-entropy loss (Eq.~\ref{eq3}; dot-dash lines). 
    The purple and yellow lines depict records for the training and validation 
    data, respectively. We started the model training without the fine-tuning 
    (FT), and then switch to that with the FT since 20 epochs (the black solid 
    vertical line). We employ the model in the 66th epoch (the black dotted 
    vertical line) that has minimum validation loss as represented by the red 
    circle. 
    }
    \label{fig4}
\end{figure}

An outline of the model architecture is presented in Fig.~\ref{fig3}. The 
classifier passes through dense layers (linear transformation), batch 
normalization \citep{Ioffe2015}, and ReLU activation ($\equiv max(0, x)$; 
\cite{Glorot2011}), and then outputs the probability value ($\hat{y}=[0:1]$) 
through a sigmoid function. 
All the HSC images are resized from $64\times64$ to $160\times160$ pixels to 
tailor the input size to the requirement of the {\tt Xception} architecture 
\citep{Chollet2016}. 
Also, for the same reason, our one-channel gray-scale data are distributed 
to all three color channels (Fig.~\ref{fig3}), and their pixel values are 
standardized in each channel by the same values used in the pre-training with the 
ImageNet \citep{Chollet2016}. 
The training data are transferred to the {\tt Xception} model with data 
augmentation built in the {\tt Keras} library ({\tt ImageDataGenerator};  
Fig.~\ref{fig3}), which performs random image rotation ($<90$ degrees), as well 
as horizontal and vertical flips.

\subsection{Model training}\label{s32}

\begin{figure}
\begin{center}
	\includegraphics[width=7.8cm]{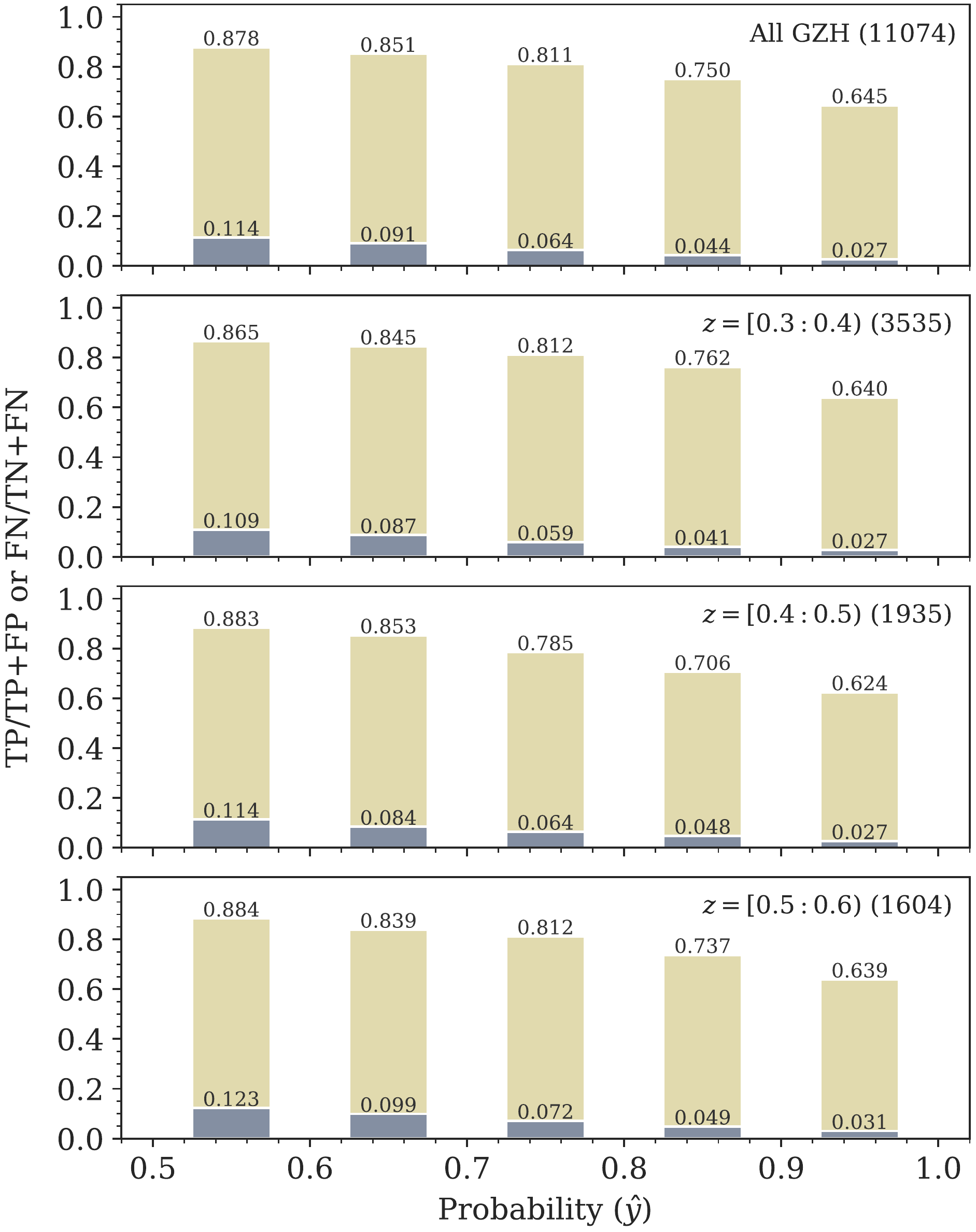}
\end{center}
    \caption{
    The precision (yellow) and false detection fraction among non-spirals (gray) 
    for the test data given the probability thresholds (the top panel) and those 
    breakdown at three redsfhit bins of $z=[0.3:0.4)$, $[0.4:0.5)$, and $[0.5:0.6)$ 
    (the 2nd--4th rows). 
    The precision is defined as the fraction of spirals correctly classified in the 
    entire spiral sample, while the false detection rate means the fraction of 
    non-spirals classified as spirals mistakingly in the non-spiral sample. 
    This work adopts $\hat{y}=0.9$ as the selection threshold of spiral galaxies.
    }
    \label{fig5}
\end{figure}

\begin{figure*}
\begin{center}
	\includegraphics[width=17cm]{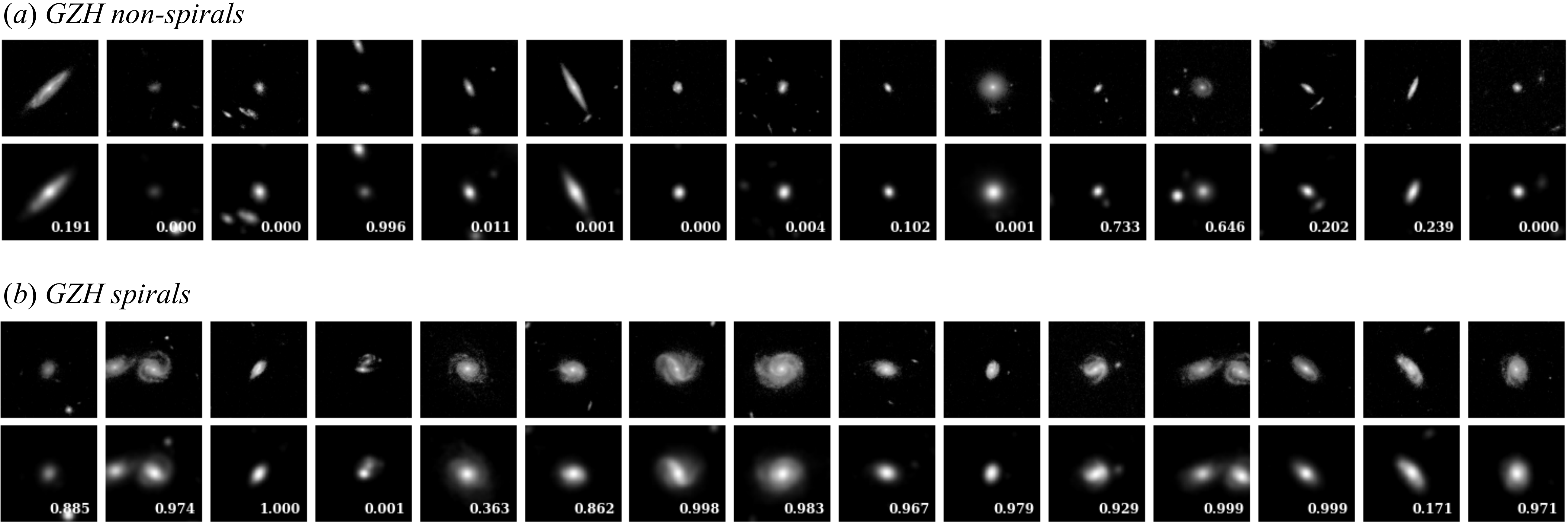}
\end{center}
    \caption{
    Comparisons with visual classifications by Galaxy Zoo Hubble (GZH; 
    \cite{Willett2017}). From the top, figures present examples of (a) non-spiral 
    and (b) spiral galaxies in GZH. 
    In each figure, upper and lower panels show the gray-scale images from the 
    COSMOS HST/ACS data \citep{Koekemoer2007} and from the HSC-SSP Wide layer 
    with arcsinh normalization (see \S\ref{s23}), respectively. 
    The probability values of spirals predicted by our deep learning model are 
    represented at the bottom-right corner in each image. 
    }
    \label{fig6}
\end{figure*}

For the model training, we employ $i$-band magnitude limited galaxies ($i<22.5$) 
from the HSC-SSP DUD layer \citep{Aihara2019}, which have counterparts to the 
Galaxy Zoo Hubble sources at $z=$ 0.2--0.7 within 1 arcsec radius 
\citep{Willett2017}. 
We construct their gray-scale cutouts in the same manner as the primary data from 
the HSC-SSP Wide layer, but the imaging depths of the DUD layer used in this work 
are by 1.5 mag deeper than those of the Wide layer. 
The Galaxy Zoo Hubble offers galaxy morphological classification conducted by 
volunteers for the images taken by the ACS camera on the Hubble Space Telescope 
(HST), which has 6.4 times better seeing FWHM (0.11 arcsec) than our 
seeing-matched HSC data (0.7 arcsec). 
They thus inform us of more reliable morphology information on the poorly-resolved 
HSC images used in this work. 
Following the selection procedure and thresholds of previous selection tasks 
suggested by \citet[section~6 and table~11]{Willett2017}, 
we select 11~074 galaxies for the model training and validation, out of which 2060 
and 9014 sources are classified as spirals and non-spirals with the final weighted 
votes $>20$. 
The spiral galaxy sample consists of 2049 normal spirals and 11 clumpy spirals 
({\tt spiral\_weighted\_fraction>0.5}), and the non-spiral sample contains 8466 
smooth objects ({\tt smooth\_best\_fraction>0.5}), 455 edge-on galaxies 
({\tt edgeon\_weighted\_fraction>0.5}), and 149 non-spirals with non-smooth 
features such as disks and clumps ({\tt no\_spiral\_weighted\_fraction>0.5}; see 
the literature for detailed information). 

By using the classification flags defined above, we performed the model training 
with the training data of spirals and non-spirals ($N=$ 1648 for each) and their 
validation data ($N=$ 412 for each). We use all spiral galaxies but randomly 
select 2060 non-spiral galaxies from their whole sample to maintain the same 
sample size as spirals. Training histories of model accuracy and binary 
cross-entropy loss are summarized in Fig.~\ref{fig4}. The model accuracy is 
defined as: 
\begin{equation}
    accuracy \equiv \frac{TP+TN}{TP+TN+FP+FN},
\label{eq2}
\end{equation}
where $TP$ (or $TN$) and $FP$ (or $FN$) represent the numbers of true positive (or 
negative) and false positive (or negative), respectively. The former refers to the 
number of predictions where the classifier correctly predicts the positive (or 
negative) class as positive (or negative), respectively. However, the latter 
refers to the number of failed predictions. The binary cross-entropy loss is 
described in the following, 
\begin{equation}
    loss \equiv -y\log\hat{y} -(1-y)\log(1-\hat{y}),
\label{eq3}
\end{equation}
where $y$ and $\hat{y}$ represent the target value $\{0, 1\}$ and the probability  
of spirals $[0:1]$, respectively. 

We trained the model 170 times, i.e. 170 epochs, with fine-tuning since the 20th 
epoch. We then selected the best set of weight parameters exhibiting the least 
cross-entropy loss in the validation data. 
In the first 20 epochs, we trained only the weights of the classification head 
using the pre-trained CNN as feature extractor, and then trained the weights of all 
the layers of the deep learning model (so-called fine-tuning technique; 
Fig.~\ref{fig4}). 
The parameter tuning was performed by {\tt Adam} optimizer, an adaptive learning 
rate optimization algorithm \citep{Kingma2014}, through the stochastic mini-batch 
training with a batch size of 32. 
The model training started with a default learning rate of $lr=1\times10^{-3}$, 
and then a learning rate decay ($lr\rightarrow2\times10^{-4}$) was applied in and 
after 50 epochs. 

Consequently, we adopt the model in the 66th epoch (Fig.~\ref{fig4}) that achieves 
92 percent accuracy in both the training and validation data. 
The histograms in Fig.~\ref{fig5} summarize precision (completeness) values and 
fractions of false detection (contamination) against the probability for the 
test data with our final model. Here we reproduced the gray-scale images of the 
Galaxy Zoo Hubble sources from the HSC-SSP Wide layers that we actually use in the 
main analysis, which amounts to 11~074 objects. We then treat them as the test 
data-set separately from the training and validation data. 
Comparison examples of HST images of the test data with those in the HSC-SSP Wide 
layer, including obtained probability values, are shown in Fig.~\ref{fig6}. 
The figure indicates that most of the test samples are successfully classified as 
labeled in the Galaxy Zoo Hubble, even with the HSC images. We also confirm no 
clear redshift dependence of the selection completeness and contamination within 
the target redshift range ($z=$ 0.3--0.6). 

For fairly selecting spiral galaxies, determining the selection cut of the 
probability value ($\hat{y}$) is a trade-off between the selection completeness 
and contamination rate, which depend on the intrinsic spiral fraction. 
When assuming a 30 percent spiral fraction, for instance, Fig.~\ref{fig5} suggests 
that we will have 20 percent of the contamination rate, while we can achieve 
nearly 90 percent completeness of the HST-based spiral galaxies. 
This work decides to use $\hat{y}=0.9$ as the selection threshold of spiral 
galaxies to reduce the contamination fraction.

\subsection{Selection of spiral galaxies}\label{s33}

\begin{figure*}
\begin{center}
	\includegraphics[width=15cm]{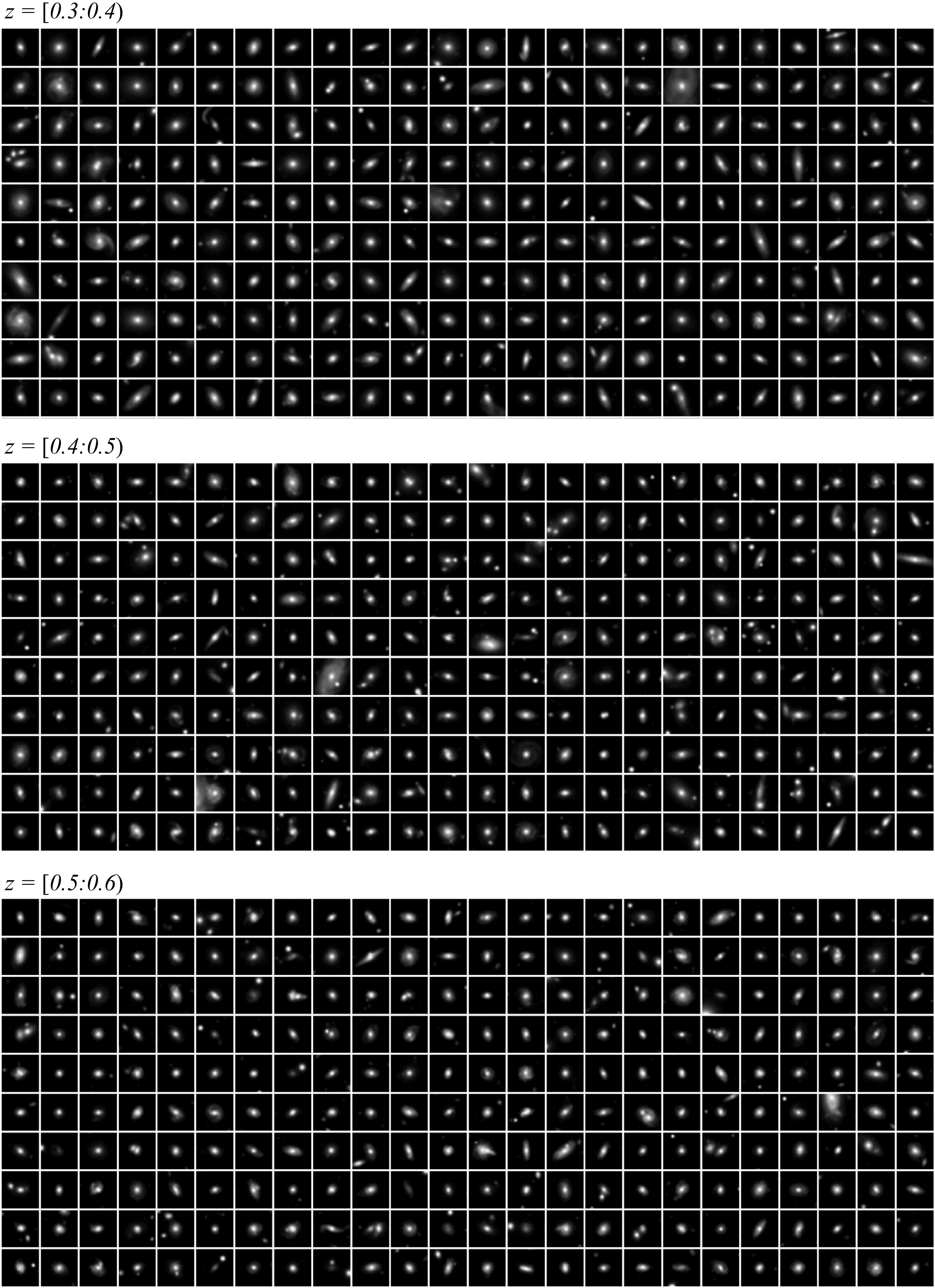}
	\vspace{5pt}
\caption{
From top to bottom, 240 random examples of gray-scale images classified as spiral 
galaxies in each redshift bin between $z=0.3$ and $z=0.6$ from the HSC-SSP Wide 
layer. 
}
\label{fig7}
\end{center}
\end{figure*}

\begin{table}
	\caption{
	Summary of the data-set. 
	The first to the fifth columns present the data for redshift bins, effective 
	volumes in  co-Mpc$^3$ (excluding mask areas), total sample size 
	(M$_\star>5\times10^{10}$ M$_\odot$), number of selected spiral galaxies, and 
	CAMIRA red sequence clusters adopted in this study \citep{Oguri2018}, 
	respectively.}
	\label{tab1}
\begin{center}
	\begin{tabular}{lrrrr} 
		\hline
		Redshift    & Volume  & Total \# & Spiral \# & RSC \#\\
		\hline
		$[0.3:0.4)$ & $5.5\times10^{7}$ &  51~192 & 16~729 & 352\\
		$[0.4:0.5)$ & $8.2\times10^{7}$ &  96~088 & 23~983 & 386\\
		$[0.5:0.6)$ & $1.1\times10^{8}$ & 128~941 & 33~391 & 398\\
		\hline
	\end{tabular}
\end{center}
\end{table}

Based on the constructed model, we search for spiral galaxies from 276~220 
stellar-mass limited samples (\S\ref{s21}). We selected galaxies with probability 
values greater than $\hat{y}>0.9$ as spiral galaxies to reduce the contaminant 
fraction (\S\ref{s32}). 
Consequently, 74~103 objects are classified as spiral galaxies, which 
represent 27 percent of the entire sample. The numbers of selected spirals and the 
effective survey volume in three redshift slices are summarized in 
Table~\ref{tab1}. Also, randomly assigned 240 postage stamps of spiral galaxies in 
each redshift bin are presented in Fig.~\ref{fig7}. 

When assuming the completeness and contamination rates inferred from the 
Galaxy Zoo Hubble (\S\ref{s32} and Fig.~\ref{fig5}), the corrected spiral 
fractions at $z=$ 0.3--0.6 are estimated to be 37--49 percent. These fractions 
broadly agree with those suggested based on the images taken from the HST/ACS 
\citep[section~4]{Tasca2009}, where $f_\mathrm{spiral}\sim$ 0.4--0.6 in the 
similar stellar mass limit. 
Thus, we consider that our deep learning model can reasonably select spiral 
galaxies in the corresponding redshift spaces, despite poorer spatial resolution of 
the HSC images than that of the HST data. 
However, we should still be missing significant fractions of spiral galaxies, 
especially at higher redshift bins, because the HST-based training and validation 
data would be incomplete. We note that this study does not aim to derive the 
intrinsic spiral fraction. Rather, we solely focus on systematic comparisons 
between the spiral fraction and the large-scale galaxy overdensity based on a 
stacking analysis, by leveraging survey volumes that are significantly larger than 
those in previous studies at the similar redshift range. 

\begin{figure}
\begin{center}
	\includegraphics[width=7.5cm]{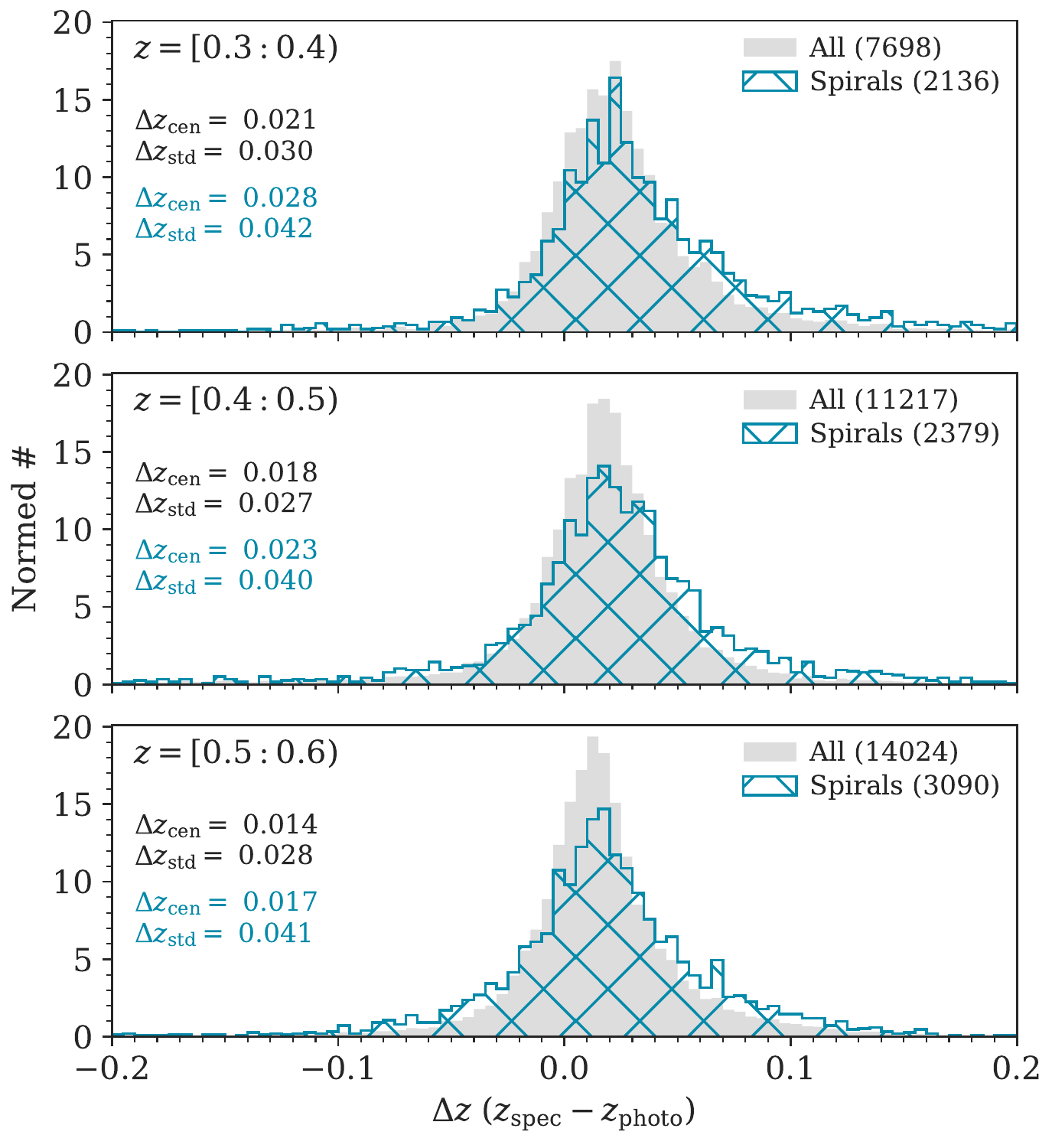}
\end{center}
    \caption{
    Normalized distributions of differentials between spectroscopic redshifts and 
    photometric redshifts ($\Delta z$) of the targets (gray) and selected spirals 
    (cyan crosshatch) with spec-$z$ at three redshift bins of $z=[0.3:0.4)$, 
    $[0.4:0.5)$, and $[0.5:0.6)$. Biweight centers ($\Delta z_\mathrm{cen}$) and 
    standard deviations ($\Delta z_\mathrm{std}$) at each redshift bin for the 
    entire sample and spiral galaxies are respectively illustrated by black and 
    cyan colors in each panel. 
    }
    \label{fig8}
\end{figure}

On top of that, we verified the precision of the photometric redshift measurement 
by {\tt Mizuki} \citep{Tanaka2015} for spiral galaxies in the target redshift 
range $z=[0.3:0.6)$. It is expected that photometric redshift errors of spiral 
galaxies would be larger than those of the entire sample because late-type galaxies 
tend to be bluer (i.e., less prominent Balmer break features) in the rest-frame 
optical wavelength. 
With this motivation, we utilize 32~939 galaxies from our sample with spectroscopic 
redshifts, out of which 7605 sources are classified as spirals, to verify the 
photo-$z$ accuracy between spirals and the total sample. 
Their spectroscopic redshifts are based on various surveys available in the HSC-SSP 
PDR2 \citep[\S5.4]{Aihara2019}: SDSS DR12--14 
\citep{Alam2015,Albareti2017,Abolfathi2018}, GAMA DR2  \citep{Liske2015}, PRIMUS 
\citep{Coil2011,Cool2013}, VIPERS \citep{Garilli2014}, VVDS \citep{Le2013}, WiggleZ 
\citep{Drinkwater2010}, and zCOSMOS \citep{Lilly2009}. 

Spectroscopic redshifts against photo-$z$ for both the entire spec-$z$ references 
and selected spirals are illustrated in Fig.~\ref{fig8}. The figure suggests that 
spiral galaxies would exhibit slightly larger photo-$z$ errors (by a factor of 
$\sim1.5$) at each redshift bin, while their redshift distributions are broadly 
consistent with each other. 
Given the adequate amount of the spiral sample at each redshift bin (N $>$ 
15~000; Table~\ref{tab1}), this study ignores such a small increase of the 
redshift uncertainty hereafter.

\section{Large-scale environmental dependence}\label{s4}

Based on the result of our deep-learning-based spiral classification and the 
density measurements from the literature \citep{Shimakawa2021,Oguri2018}, we 
investigate whether or not there is a large-scale morphology--density relation 
(\S\ref{s41}). 
We also study distributions of the rest-frame $u-r$ colors of spirals in different 
density environments (\S\ref{s42}).

\subsection{Spiral fractions across large-scale environments}\label{s41}

\begin{figure*}
\begin{center}
\includegraphics[width=15cm]{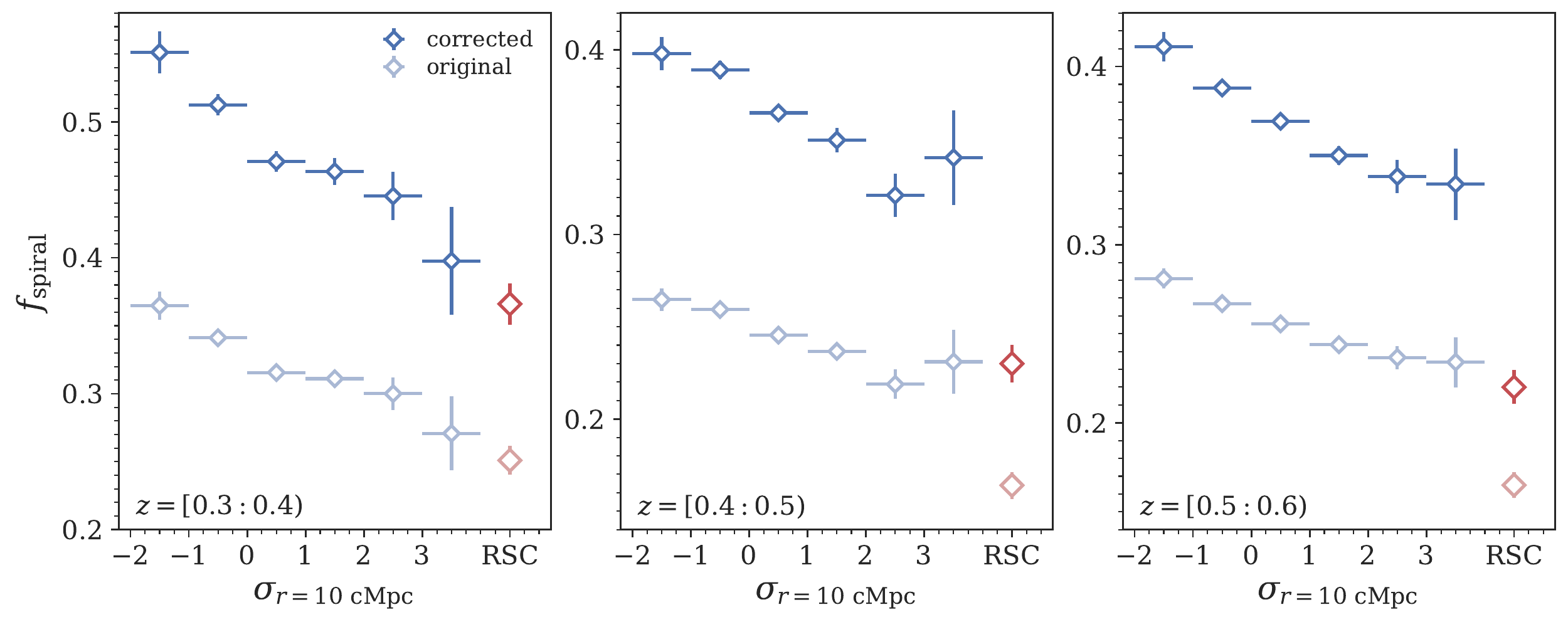}
\end{center}
\caption{
    From left to right, spiral fractions as a function of the over-density at 
    redshift bins of $z=[0.3:0.4)$, $[0.4:0.5)$, and $[0.5:0.6)$, respectively. 
    The red symbol in the far right in each panel depicts the spiral fraction in 
    the vicinity ($<2$ co-Mpc) of red sequence clusters (RSC) from 
    \citet{Oguri2018}. Vertical error bars indicate Poisson errors. 
    The light color symbols are based on direct measurements with the selection 
    threshold of the probability $\hat{y}=0.9$, and the dark color symbols are 
    scaled from the original values based on the completeness and contamination 
    rates of the spiral classification inferred from the test data (see \S\ref{s32} 
    and Fig.~\ref{fig5}). 
    }
    \label{fig9}
\end{figure*}

With 74~103 spiral galaxies out of 276~220 stellar-mass limited samples 
(Table~\ref{tab1}), we estimate the spiral fractions in different overdensities 
and red sequence clusters from the archive catalog 
(\cite[see Section 2.2]{Shimakawa2021,Oguri2018}), at three target redshift bins, 
$z=[0.3:0.4)$, $[0.4:0.5)$, and $[0.5:0.6)$. 
The result is represented in Fig.~\ref{fig9}, where we stacked all sources and 
spirals given the similar overdensity $\Delta\sigma_\mathrm{r=10cMpc}=1$ across the 
entire survey field to accumulate the sufficient sample size in each bin. 
The red diamonds also indicate the spiral fractions in the vicinity of red sequence 
clusters selected by the CAMIRA cluster catalog within the same survey field 
\citep{Oguri2018} at the same redshift bin in each panel. 
Typical overdensities ($\sigma_\mathrm{r=10cMpc}$) of red-sequence clusters in 
\citet{Shimakawa2021} are $\sigma_\mathrm{r=10cMpc}=1.1$, $0.9$, and $1.1$ at 
$z=[0.3:0.4)$, $[0.4:0.5)$, and $[0.5:0.6)$, respectively. 
We selected galaxies in 2 co-Mpc projected distances from the red sequence clusters. 
The distance threshold was decided considering that the excess of red fractions has 
been detected within $\lesssim2$ co-Mpc around the CAMIRA clusters 
\citep[\S5.2]{Nishizawa2018}. 
Fig.~\ref{fig9} shows both the spiral fractions from direct measurements and those 
with corrections of the completeness and contamination rates given in 
Fig.~\ref{fig5}. The contamination rates within the selected spirals are estimated 
to be approximately 5--8 percent. 

Owing to the combination of the wide-field deep imaging data by HSC-SSP and the 
deep learning classification, Fig.~\ref{fig9} successfully demonstrates moderate 
trends of the large-scale environmental dependence of the spiral fraction at 
$z=$ 0.3--0.6 related to lower spiral fractions at higher overdensities on 
10 co-Mpc scale. 
What is worth noting is that we find a systematic increase of the spiral fraction 
in lower-density environments at each redshift slice, which is achievable only with 
such continuous overdensity information by our previous work \citep{Shimakawa2021}. 
Moreover, we confirmed deficits of the spiral fraction in the vicinity of red 
sequence clusters at all three redshift bins, as reported in previous studies 
(e.g., \cite{Moran2007,Poggianti2009} and references therein).

\subsection{Rest-frame colors of spirals vs environments}\label{s42}

Then, we investigated the color dependence of detected spiral galaxies on the 
galaxy overdensity. Their rest-frame $u-r$ colors are derived by the SED-based 
fitting code, {\tt Mizuki} \citep{Tanaka2015}. This is motivated by the well-known 
environmental dependence of the red fraction (e.g., 
\cite{Tanaka2005,Peng2010,Alpaslan2015}). In particular, many previous studies have 
reported that passive spiral galaxies (so-called anemic spirals; \cite{Bergh1976}) 
are more frequently observed in galaxy clusters or intermediate overdensities 
(e.g. \cite{Goto2003a,Boselli2006,Blanton2009,Bamford2009,Masters2010} and 
references therein). 
Similar to those already observed in the past, there could be a color dependency of 
spiral galaxies even on large-scale environments. 

\begin{figure*}
\begin{center}
	\includegraphics[width=15cm]{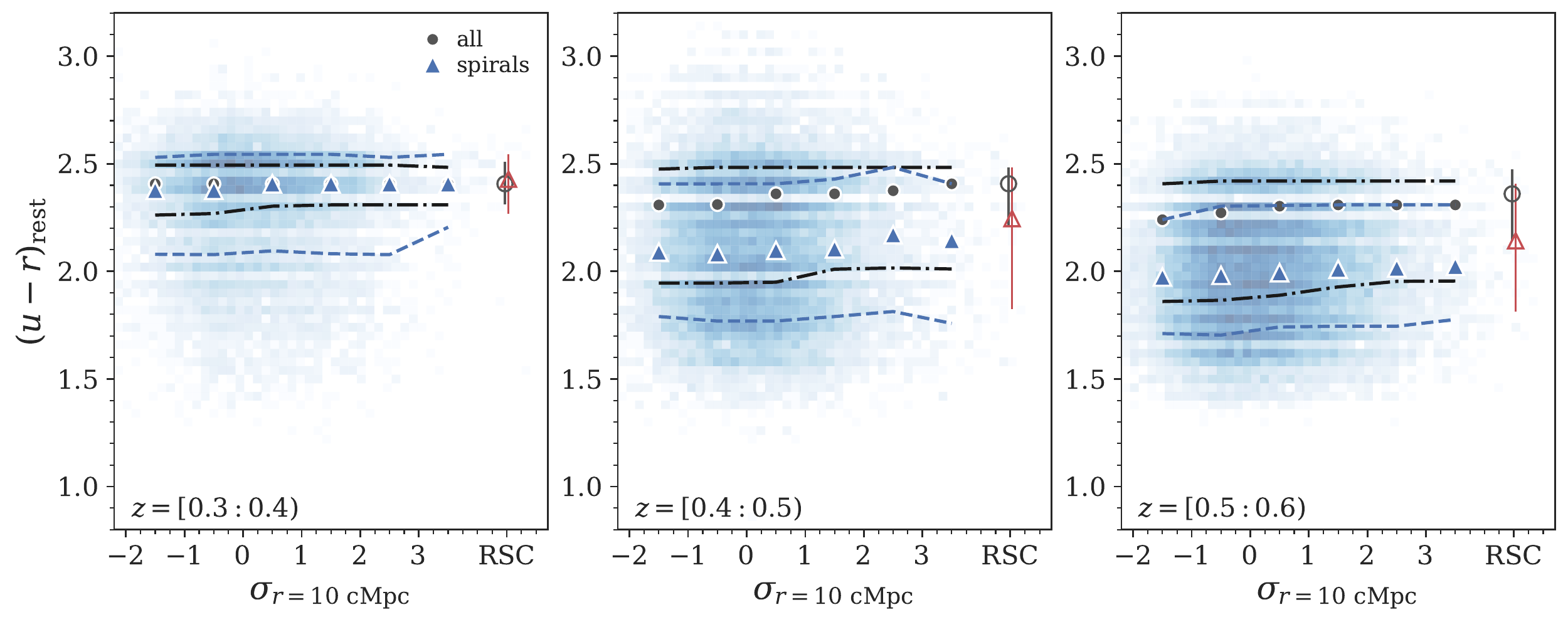}
\end{center}
    \caption{
    Distributions of the rest-frame $u-r$ colors of the entire sample and spiral 
    galaxies as a function of the galaxy over-density ($\sigma_\mathrm{r=10cMpc}$). 
    Black-filled circles and dot-dash lines represent the median and 68th 
    percentiles for the entire source, and blue triangles and dash lines show 
    those for the spirals. The black open circles and red open triangles are 
    massive galaxies and spirals nearby red sequence clusters \citep{Oguri2018}. 
    The error bars indicate 68th percentile regions. 
    We note that the similar color distributions of spirals to the entire sample 
    at $z=$ 0.3--0.4 should be due to the measurement error because their 
    rest-frame $u$-band fluxes are not fully covered by the HSC $grizy$ filters. 
    }
    \label{fig10}
\end{figure*}

The derived rest-frame $u-r$ color distributions of spiral galaxies as a function 
of the overdensity are illustrated in Fig.~\ref{fig10}. Dashed lines and red error 
bars represent 68th percentiles of spiral galaxies given overdensity and red 
sequence clusters, respectively. 
We observe bluer color distributions of spiral galaxies compared to those in the 
entire sample except the lowest-$z$ sample at $z=[0.3:0.4)$. However, the similar 
colors between spirals and all sources at $z=[0.3:0.4)$ should be due to a 
technical issue since the HSC filters ($grizy$) do not fully cover the rest-frame 
$u$-band photometry of galaxies at this redshift range. 
Over the redshift range of $z=$ 0.3--0.6, in general, we cannot determine any 
significant color dependence on the large-scale over-density given the spiral 
sample, while there is a lack of relatively bluer spirals in the highest density 
bin and clusters at $z=[0.3:0.4)$ (Fig.~\ref{fig10}). 
A possible explanation of a decorrelation between rest-frame $u-r$ colors of 
spirals and the galaxy overdensities is described in the discussion section 
(\S\ref{s5}). 
On the other hand, median $u-r$ colors of spirals in the vicinity of red sequence 
clusters are slightly redder than those in the other fields at $z=[0.4:0.5)$ and 
$z=[0.5:0.6)$, which broadly agrees with the observational trends reported in the 
previous work \citep{Goto2004,Wolf2007,Bamford2009,Cantale2016}.

\section{Discussion}\label{s5}

We discuss and attempt to interpret the results obtained in the last 
section. Based on the stacking analysis, we determined a statistical trend in 
which the spiral fraction tends to increase (or decrease) in lower (or higher)  
densities at $z=$ 0.3--0.6 on ten co-Mpc scales. 
This suggests a bias on the global scale across the large-scale structure of the 
universe, which could originate from a difference in mass assembly histories in the 
underlying density environments of the large-scale structure. In fact, it is 
plausible that the known morphology--density relation in and around massive 
clusters can be validated even on such a wide scale, because large-scale 
overdensities tend to be associated with larger numbers of clusters, as 
demonstrated in the study by \citet{Shimakawa2021}. 

\begin{figure}
\begin{center}
	\includegraphics[width=7.5cm]{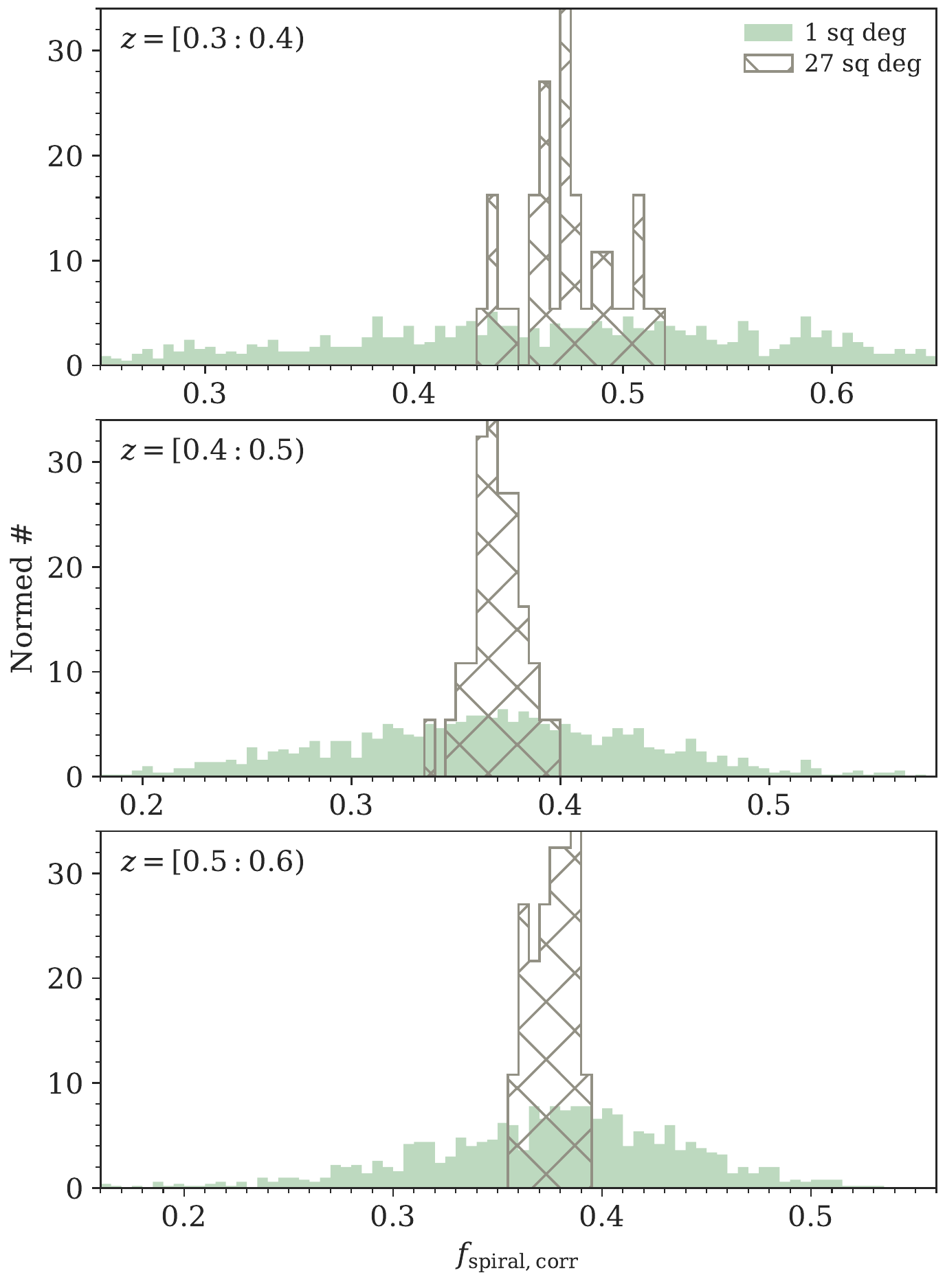}
\end{center}
    \caption{
    Normalized distributions of the corrected spiral fractions in random 1 deg$^2$ 
    (light green filled) and 27 deg$^2$ (dark green crosshatch) fields at 
    $z=[0.3:0.4)$, $[0.4:0.5)$, and $[0.5:0.6)$. 
    }
    \label{fig11}
\end{figure}

Moderate environmental dependence of galaxy morphology on the large-scale will lead 
to a non-negligible cosmic variance in the measurement of the spiral fraction if 
research does not use the data covering sufficiently large areas. 
For testing such a cosmic variance effect, we investigate the variation of the 
spiral fraction in individual fields given survey areas at each redshift bin. 
Fig.~\ref{fig11} represents the variance of the spiral fractions 
($f_\mathrm{spiral}$) within 1 and 27 deg$^2$ based on 1000 random points in our 
survey field, where 27 deg$^2$ corresponds to the footprint of the HSC-SSP DUD 
layer. The figure suggests a significant cosmic variance in the derivation of the 
spiral fraction in 1 deg$^2$. However, measurement errors due to the small sample 
size ($N=$ 182, 341, and 457 per deg$^2$ on average at $z=[0.3:0.4)$, $[0.4:0.5)$, 
and $[0.5:0.6)$) would contribute to some increase of the variance. 
Even if the survey area reaches 27 deg$^2$, there are small variations of 
$\pm0.02$ to $0.04$ depending on the redshift. 
The significant variability also cautions that, when we study morphology--density 
dependence on the local scale, we may need to take care of the underlying 
large-scale structure of the target region. Such comprehensive work requires an 
expensive spectroscopic survey with extremely wide-field coverage, such as 
DESI \citep{DESI2016}, PFS \citep{Takada2014,Tamura2016}, and 4MOST 
\citep{de2012,de2019}. These programs will allow us to unveil a systematic bias of 
the spiral fraction in various environments on a large scale as well as the local 
scale. 

Next, selected spiral objects do not exhibit a clear environmental dependence on 
the rest-frame $u-r$ color. Previous studies have reported that galaxy clusters 
and/or groups tend to have a higher fraction of red spirals 
\citep{Goto2003b,Wolf2007,Bamford2009,Masters2010}, thus suggesting a transition of 
spiral galaxies falling into massive halos to S0 galaxies, as observed in today's 
massive clusters \citep{Goto2004,Moran2007,Kovac2010}. We reconfirm such a 
systematic color trend in the vicinity of red sequence clusters in our sample. 
However, regarding large-scale environments (10 co-Mpc in our case), such a 
physical transition occurring on the local scale ($\sim$ a physical Mpc) becomes 
negligible, which might explain the absence of a significant trend in this study. 

We should stress that the current spiral classification is highly limited to bright 
and massive objects with a limiting stellar mass of $5\times10^{10}$ M$_\odot$.  
Also, we would miss a non-negligible fraction of spiral galaxies owing to the 
resolution limit (0.7 arcsec in seeing FWHM), though we used the HST-based 
morphological information as classification flags by \citet{Willett2017}. 
Such a selection bias is a major challenge that needs to be addressed in the 
future, as stated at the end.

\section{Conclusions}\label{s6}

Lastly, we summarize the flow of this paper. We investigated the spiral fraction of 
mass-limited galaxies based on the sample in the significant volumes obtained 
from the HSC-SSP PDR2. This study attempted to elucidate the global-scale 
environmental dependence that may originate from the varying growth of underlying 
density environments for the large-scale structure rather than physical 
interactions between galaxies and clusters occurring on the local scale. Previous 
studies, such as the SED fitting \citep{Tanaka2018}, projected density map 
\citep{Shimakawa2021}, and red sequence cluster search \citep{Oguri2018} 
based on HSC-SSP, enabled us to achieve such an objective. 
Consequently, we obtained 276~220 massive galaxies across 360 deg$^2$, parts of 
which are associated with 1136 red sequence clusters at the corresponding redshift 
range. A total of 74~103 galaxies were labeled as spiral galaxies via transfer 
learning based on a pre-trained deep CNN. The obtained results indicate the 
moderate large-scale environmental dependence of the spiral fraction at 
$z=$ 0.3--0.6. 
Furthermore, we do not observe any obvious trend of the rest-frame $u-r$ colors for 
the spiral galaxies towards the overdensity. 
These results suggest a large-scale environmental dependence of galaxy morphology 
at least up to $z=0.6$, while observed color transitions of spiral galaxies falling 
into galaxy clusters and groups are no longer detected on such a wide scale. 

Although the current analysis is limited by the photo-$z$ uncertainty and high 
stellar mass ($>5\times10^{10}$ M$_\odot$), ongoing and upcoming intensive 
programs will resolve these limitations. For instance, DESI \citep{DESI2016} will 
address more than a thousand spectroscopic data per square degree for 
emission-line galaxies at intermediate redshift ($z\lesssim1.3$) over $14k$ 
deg$^2$. Other forthcoming spectroscopic surveys such as PFS 
\citep{Takada2014,Tamura2016}, MOONS \citep{Cirasuolo2011,Cirasuolo2014}, and 
4MOST \citep{de2012,de2019} will also facilitate the accumulation of more detailed 
redshift information. In addition, wide-field high-resolution imaging by the Roman 
Space Telescope \citep{Spergel2015,Akeson2019} will provide a clear path for the 
morphology classification in the distant universe, which can considerably extend 
the sample size towards lower stellar mass and higher redshift from this study.


\begin{ack}
This study was conducted based on data collected at the Subaru Telescope and retrieved from the HSC data archive system, which is operated by Subaru Telescope and Astronomy Data Center at National Astronomical Observatory of Japan. We are honored and grateful for the opportunity of observing the Universe from Maunakea, which has cultural, historical, and natural significance in Hawaii. 
All the data underlying this publication are available under the Second Public Data 
Release of Hyper Suprime-Cam Subaru Strategic Program 
(\url{https://hsc.mtk.nao.ac.jp/ssp/data-release/})

The Hyper Suprime-Cam (HSC) collaboration includes the astronomical communities of Japan and Taiwan, and Princeton University. The HSC instrumentation and software were developed by the National Astronomical Observatory of Japan (NAOJ), the Kavli Institute for the Physics and Mathematics of the Universe (Kavli IPMU), the University of Tokyo, the High Energy Accelerator Research Organization (KEK), the Academia Sinica Institute for Astronomy and Astrophysics in Taiwan (ASIAA), and Princeton University. Funding was contributed by the FIRST program from Japanese Cabinet Office, the Ministry of Education, Culture, Sports, Science and Technology (MEXT), the Japan Society for the Promotion of Science (JSPS), Japan Science and Technology Agency (JST), the Toray Science Foundation, NAOJ, Kavli IPMU, KEK, ASIAA, and Princeton University. 

This paper makes use of software developed for the Large Synoptic Survey Telescope. We thank the LSST Project for making their code available as free software at  \url{http://dm.lsst.org}

The Pan-STARRS1 Surveys (PS1) have been made possible through contributions of the Institute for Astronomy, the University of Hawaii, the Pan-STARRS Project Office, the Max-Planck Society and its participating institutes, the Max Planck Institute for Astronomy, Heidelberg and the Max Planck Institute for Extraterrestrial Physics, Garching, The Johns Hopkins University, Durham University, the University of Edinburgh, Queen’s University Belfast, the Harvard-Smithsonian Center for Astrophysics, the Las Cumbres Observatory Global Telescope Network Incorporated, the National Central University of Taiwan, the Space Telescope Science Institute, the National Aeronautics and Space Administration under Grant No. NNX08AR22G issued through the Planetary Science Division of the NASA Science Mission Directorate, the National Science Foundation under Grant No. AST-1238877, the University of Maryland, and Eotvos Lorand University (ELTE) and the Los Alamos National Laboratory.

Funding for the Sloan Digital Sky Survey IV has been provided by the Alfred P. Sloan Foundation, the U.S. Department of Energy Office of Science, and the Participating Institutions. SDSS-IV acknowledges support and resources from the Center for High Performance Computing  at the University of Utah. The SDSS website is \url{www.sdss.org}.

We thank anonymous referee for helpful feedback.
We would like to thank Editage (\url{www.editage.com}) for English language editing.
This work was partially supported by Summer Student Program (2020) National Astronomical Observatory of Japan and the Department of Astronomical Science, The Graduate University for Advanced Studies, SOKENDAI.

This study made extensive use of the following tools, {\tt NumPy} 
\citep{Harris2020}, the Tool for OPerations on Catalogues And Tables, {\tt TOPCAT} 
\citep{Taylor2005}, a community-developed core Python package for Astronomy, 
{\tt Astropy} \citep{Astropy2013}, and Python Data Analysis Library {\tt pandas} 
\citep{McKinney2010}. 
\end{ack}




\begin{thebibliography}{}
\bibitem[Abadi et al.(2016)]{Abadi2016}
  Abadi, M., et al.\ 2016, arXiv e-prints, arXiv:1603.04467
\bibitem[Abazajian et al.(2004)]{Abazajian2004}
  Abazajian K., et al.\ 2004, AJ, 128, 502
\bibitem[Abolfathi et al.(2018)]{Abolfathi2018}
  Abolfathi, B., et al.\ 2018, \apjs, 235, 42
\bibitem[Ackermann et al.(2018)]{Ackermann2018}
  Ackermann, S., et al.\ 2018, \mnras, 479, 415
\bibitem[Aihara et al.(2018)]{Aihara2018}
  Aihara, H., et al.\ 2018, \pasj, 70, S4
\bibitem[Aihara et al.(2019)]{Aihara2019}
  Aihara, H., et al.\ 2019, \pasj, 71, 114
\bibitem[Akeson et al.(2019)]{Akeson2019}
  Akeson, R., et al.\ 2019, arXiv e-prints, arXiv:1902.05569
\bibitem[Alam et al.(2015)]{Alam2015}
  Alam, S., et al.\ 2015, \apjs, 219, 12
\bibitem[Albareti et al.(2017)]{Albareti2017}
  Albareti, F.~D., et al.\ 2017, \apjs, 233, 25
\bibitem[Alpaslan et al.(2015)]{Alpaslan2015}
  Alpaslan, M., et al.\ 2015, \mnras, 451, 3249
\bibitem[Astropy Collaboration et al.(2013)]{Astropy2013}
  Astropy Collaboration, Robitaille, et al.\ 2013, \aap, 558, A33
\bibitem[Baldry et al.(2006)]{Baldry2006}
  Baldry, I.~K., et al.\ 2006, \mnras, 373, 469
\bibitem[Balogh et al.(2004)]{Balogh2004}
  Balogh, M.~L., et al.\ 2004, \apjl, 615, L101
\bibitem[Bamford et al.(2009)]{Bamford2009}
  Bamford, S.~P., et al.\ 2009, \mnras, 393, 1324
\bibitem[Banerji et al.(2010)]{Banerji2010}
  Banerji, M., et al.\ 2010, \mnras, 406, 342
\bibitem[Bekki and Couch(2011)]{Bekki2011}
  Bekki, K., \& Couch, W.~J.\ 2011, \mnras, 415, 1783
\bibitem[Blanton and Moustakas(2009)]{Blanton2009}
  Blanton, M.~R., \& Moustakas, J.\ 2009, \araa, 47, 159
\bibitem[Bosch et al.(2018)]{Bosch2018}
  Bosch, J., et al.\ 2018, \pasj, 70, S5
\bibitem[Boselli and Gavazzi(2006)]{Boselli2006}
  Boselli, A., \& Gavazzi, G.\ 2006, \pasp, 118, 517
\bibitem[Bower et al.(1992)]{Bower1992}
  Bower, R.~G., Lucey, J.~R., \& Ellis, R.~S.\ 1992, \mnras, 254, 601
\bibitem[Bower et al.(1998)]{Bower1998}
  Bower, R.~G., Kodama, T., \& Terlevich, A.\ 1998, \mnras, 299, 1193
\bibitem[Bozinovski and Fulgosi(1976)]{Bozinovski1976}
  Bozinovski, S., \& Fulgosi, A.\ 1976, Proc. Symp. Informatica 3-121-5
\bibitem[Brough et al.(2017)]{Brough2017}
  Brough, S., et al.\ 2017, \apj, 844, 59
\bibitem[Butcher and Oemler(1984)]{Butcher1984}
  Butcher, H., \& Oemler, A.\ 1984, \apj, 285, 426
\bibitem[Cantale et al.(2016)]{Cantale2016}
  Cantale, N., et al.\ 2016, \aap, 589, A82
\bibitem[Capak et al.(2007)]{Capak2007}
  Capak, P., et al.\ 2007, \apjs, 172, 284
\bibitem[Cappellari et al.(2011)]{Cappellari2011}
  Cappellari, M., et al.\ 2011, \mnras, 416, 1680
\bibitem[Chabrier(2003)]{Chabrier2003}
  Chabrier, G.\ 2003, \pasp, 115, 763
\bibitem[Cheng et al.(2020)]{Cheng2020}
  Cheng, T.-Y., et al.\ 2020, \mnras, 493, 4209
\bibitem[Chollet et al.(2015)]{Chollet2015}
  Chollet, F., et al.\ 2015, Available at: \url{https://keras.io}
\bibitem[Chollet(2016)]{Chollet2016}
  Chollet, F.\ 2016, arXiv e-prints, arXiv:1610.02357
\bibitem[Cirasuolo et al.(2011)]{Cirasuolo2011}
  Cirasuolo, M., et al.\ 2011, The Messenger, 145, 11
\bibitem[Cirasuolo et al.(2014)]{Cirasuolo2014}
  Cirasuolo, M., et al.\ 2014, \procspie, 9147, 91470N
\bibitem[Coil et al.(2011)]{Coil2011}
  Coil, A.~L., et al.\ 2011, \apj, 741, 8
\bibitem[Cool et al.(2013)]{Cool2013}
  Cool, R.~J., et al.\ 2013, \apj, 767, 118
\bibitem[Cooper et al.(2007)]{Cooper2007}
  Cooper, M.~C., et al.\ 2007, \mnras, 376, 1445
\bibitem[Costa-Duarte et al.(2018)]{Costa-Duarte2018}
  Costa-Duarte, M.~V., et al.\ 2018, \mnras, 478, 1968
\bibitem[Couch et al.(1998)]{Couch1998}
  Couch, W.~J., et al.\ 1998, \apj, 497, 188
\bibitem[Coupon et al.(2018)]{Coupon2018}
  Coupon, J., et al.\ 2018, \pasj, 70, S7
\bibitem[Davidzon et al.(2017)]{Davidzon2017}
  Davidzon, I., et al.\ 2017, \aap, 605, A70
\bibitem[Davis et al.(1985)]{Davis1985}
  Davis, M., et al.\ 1985, \apj, 292, 371
\bibitem[de Jong et al.(2012)]{de2012}
  de Jong, R.~S., et al.\ 2012, \procspie, 8446, 84460T
\bibitem[de Jong et al.(2019)]{de2019}
  de Jong, R.~S., et al.\ 2019, The Messenger, 175, 3
\bibitem[de Vaucouleurs(1961)]{Vaucouleurs1961}
  de Vaucouleurs, G.\ 1961, \apjs, 5, 233
\bibitem[Dekel and Silk(1986)]{Dekel1986}
  Dekel, A., \& Silk, J.\ 1986, \apj, 303, 39
\bibitem[DESI Collaboration et al.(2016)]{DESI2016}
  DESI Collaboration, Aghamousa, et al.\ 2016, arXiv e-prints, arXiv:1611.00036
\bibitem[Dieleman et al.(2015)]{Dieleman2015}
  Dieleman, S., Willett, K.~W., \& Dambre, J.\ 2015, \mnras, 450, 1441
\bibitem[Dom{\'\i}nguez S{\'a}nchez et al.(2019)]{Dominguez2019}
  Dom{\'\i}nguez S{\'a}nchez, H., et al.\ 2019, \mnras, 484, 93
\bibitem[Donnari et al.(2021)]{Donnari2021}
  Donnari, M., et al.\ 2021, \mnras, 500, 4004
\bibitem[Dressler(1980)]{Dressler1980}
  Dressler, A.\ 1980, \apj, 236, 351
\bibitem[Dressler et al.(1997)]{Dressler1997}
  Dressler, A., et al.\ 1997, \apj, 490, 577
\bibitem[Drinkwater et al.(2010)]{Drinkwater2010}
  Drinkwater, M.~J., et al.\ 2010, \mnras, 401, 1429
\bibitem[Fogarty et al.(2014)]{Fogarty2014}
  Fogarty, L.~M.~R., et al.\ 2014, \mnras, 443, 485
\bibitem[Furusawa et al.(2018)]{Furusawa2018}
  Furusawa, H., et al.\ 2018, \pasj, 70, S3
\bibitem[Garilli et al.(2014)]{Garilli2014}
  Garilli, B., et al.\ 2014, \aap, 562, A23
\bibitem[Ghosh et al.(2020)]{Ghosh2020}
  Ghosh, A., et al.\ 2020, \apj, 895, 112
\bibitem[Gladders and Yee(2005)]{Gladders2005}
  Gladders, M.~D., \& Yee, H.~K.~C.\ 2005, \apjs, 157, 1
\bibitem[Glorot et al.(2011)]{Glorot2011}
  Glorot, X., Bordes, A., \& Bengio. Y.\ 2011, In Proc. 14th International Conf. on Artificial Intelligence and Statistics, 315, 323
\bibitem[Goto et al.(2003a)]{Goto2003a}
  Goto, T., et al.\ 2003, \mnras, 346, 601
\bibitem[Goto et al.(2003b)]{Goto2003b}
  Goto, T., et al.\ 2003, \pasj, 55, 757
\bibitem[Goto et al.(2004)]{Goto2004}
  Goto, T., et al.\ 2004, \mnras, 348, 515
\bibitem[Goto(2005)]{Goto2005}
  Goto, T.\ 2005, \mnras, 356, L6
\bibitem[Greene et al.(2017)]{Greene2017}
  Greene, J.~E., et al.\ 2017, \apjl, 851, L33
\bibitem[Gr{\"u}tzbauch et al.(2011)]{Grutzbauch2011}
  Gr{\"u}tzbauch, R., et al.\ 2011, \mnras, 411, 929
\bibitem[Guzzo et al.(2007)]{Guzzo2007}
  Guzzo, L., et al.\ 2007, \apjs, 172, 254
\bibitem[Harris et al.(2020)]{Harris2020}
  Harris, C.~R., et al.\ 2020, \nat, 585, 357
\bibitem[Hashimoto et al.(2019)]{Hashimoto2019}
  Hashimoto, T., et al.\ 2019, \mnras, 489, 2014
\bibitem[Hinshaw et al.(2013)]{Hinshaw2013}
  Hinshaw, G., et al.\ 2013, \apjs, 208, 19
\bibitem[Holden et al.(2007)]{Holden2007}
  Holden, B.~P., et al.\ 2007, \apj, 670, 190
\bibitem[Houghton et al.(2013)]{Houghton2013}
  Houghton, R.~C.~W., et al.\ 2013, \mnras, 436, 19
\bibitem[Hunter(2007)]{Hunter2007}
  Hunter, J.~D.\ 2007, Computing in Science and Engineering, 9, 90
\bibitem[Ilbert et al.(2010)]{Ilbert2010}
  Ilbert, O., et al.\ 2010, \apj, 709, 644
\bibitem[Ioffe and Szegedy(2015)]{Ioffe2015}
  Ioffe, S., \& Szegedy, C.\ 2015, arXiv e-prints, arXiv:1502.03167
\bibitem[Joshi et al.(2020)]{Joshi2020}
  Joshi, G.~D., et al.\ 2020, \mnras, 496, 2673
\bibitem[Kauffmann et al.(2004)]{Kauffmann2004}
  Kauffmann, G., et al.\ 2004, \mnras, 353, 713
\bibitem[Kawanomoto et al.(2018)]{Kawanomoto2018}
  Kawanomoto, S., et al.\ 2018, \pasj, 70, 66
\bibitem[Kere{\v{s}} et al.(2005)]{Keres2005}
  Kere{\v{s}}, D., et al.\ 2005, \mnras, 363, 2
\bibitem[Kingma and Ba(2014)]{Kingma2014}
  Kingma, D.~P., \& Ba, J.\ 2014, arXiv e-prints, arXiv:1412.6980
\bibitem[Kodama et al.(2007)]{Kodama2007}
  Kodama, T., et al.\ 2007, \mnras, 377, 1717
\bibitem[Koekemoer et al.(2007)]{Koekemoer2007}
  Koekemoer, A.~M., et al.\ 2007, \apjs, 172, 196
\bibitem[Komiyama et al.(2018)]{Komiyama2018}
  Komiyama, Y., et al.\ 2018, \pasj, 70, S2
\bibitem[Koulouridis et al.(2021)]{Koulouridis2021}
  Koulouridis, E., et al.\ 2021, \aap, 652, A12
\bibitem[Kova{\v{c}} et al.(2010)]{Kovac2010}
  Kova{\v{c}}, K., et al.\ 2010, \apj, 718, 86
\bibitem[Kuminski and Shamir(2016)]{Kuminski2016}
  Kuminski, E., \& Shamir, L.\ 2016, \apjs, 223, 20
\bibitem[Laigle et al.(2016)]{Laigle2016}
  Laigle, C., et al.\ 2016, \apjs, 224, 24
\bibitem[Larson et al.(1980)]{Larson1980}
  Larson, R.~B., Tinsley, B.~M., \& Caldwell, C.~N.\ 1980, \apj, 237, 692
\bibitem[Lecun et al.(1998)]{Lecun1998}
  Lecun Y., et al.\ 1998, Proc. IEEE, 86, 2278
\bibitem[Lecun et al.(2015)]{Lecun2015}
  Lecun, Y., Bengio, Y., \& Hinton, G.\ 2015, \nat, 521, 436
\bibitem[Le F{\`e}vre et al.(2013)]{Le2013}
  Le F{\`e}vre, O., et al.\ 2013, \aap, 559, A14
\bibitem[Lietzen et al.(2012)]{Lietzen2012}
  Lietzen, H., et al.\ 2012, \aap, 545, A104
\bibitem[Lilly et al.(2009)]{Lilly2009}
  Lilly, S.~J., et al.\ 2009, \apjs, 184, 218
\bibitem[Liske et al.(2015)]{Liske2015}
  Liske, J., et al.\ 2015, \mnras, 452, 2087
\bibitem[Lupton et al.(1999)]{Lupton1999}
  Lupton, R.~H., Gunn, J.~E., \& Szalay, A.~S.\ 1999, \aj, 118, 1406
\bibitem[Lupton et al.(2004)]{Lupton2004}
  Lupton, R., et al.\ 2004, \pasp, 116, 133
\bibitem[Masters et al.(2010)]{Masters2010}
  Masters, K.~L., et al.\ 2010, \mnras, 405, 783
\bibitem[Maturi et al.(2019)]{Maturi2019}
  Maturi, M., et al.\ 2019, \mnras, 485, 498
\bibitem[McKinney et al.(2010)]{McKinney2010}
  McKinney W., et al.\ 2010, Proc. 9th Python Sci. Conf., 1697900, 51
\bibitem[Mei et al.(2009)]{Mei2009}
  Mei, S., et al.\ 2009, \apj, 690, 42
\bibitem[Miyazaki et al.(2018)]{Miyazaki2018}
  Miyazaki, S., et al.\ 2018, \pasj, 70, S1
\bibitem[Moore et al.(1998)]{Moore1998}
  Moore, B., Lake, G., \& Katz, N.\ 1998, \apj, 495, 139
\bibitem[Moran et al.(2007)]{Moran2007}
  Moran, S.~M., et al.\ 2007, \apj, 671, 1503
\bibitem[Muzzin et al.(2008)]{Muzzin2008}
  Muzzin, A., et al.\ 2008, \apj, 686, 966
\bibitem[Muzzin et al.(2014)]{Muzzin2014}
  Muzzin, A., et al.\ 2014, \apj, 796, 65
\bibitem[Nishizawa et al.(2018)]{Nishizawa2018}
  Nishizawa, A.~J., et al.\ 2018, \pasj, 70, S24
\bibitem[Oguri(2014)]{Oguri2014}
  Oguri, M.\ 2014, \mnras, 444, 147
\bibitem[Oguri et al.(2018)]{Oguri2018}
  Oguri, M., et al.\ 2018, \pasj, 70, S20
\bibitem[Oke and Gunn(1983)]{Oke1983}
  Oke, J.~B., \& Gunn, J.~E.\ 1983, \apj, 266, 713
\bibitem[Old et al.(2020)]{Old2020}
  Old, L.~J., et al.\ 2020, \mnras, 493, 5987
\bibitem[Peng et al.(2010)]{Peng2010}
  Peng, Y.-. jie ., et al.\ 2010, \apj, 721, 193
\bibitem[Peng et al.(2012)]{Peng2012}
  Peng, Y.-. jie ., et al.\ 2012, \apj, 757, 4
\bibitem[Pietrowski et al.(2021)]{Pietrowski2021}
  Pietrowski, M., et al.\ 2021, arXiv e-prints, arXiv:2109.01451
\bibitem[Poggianti et al.(2006)]{Poggianti2006}
  Poggianti, B.~M., et al.\ 2006, \apj, 642, 188
\bibitem[Poggianti et al.(2009)]{Poggianti2009}
  Poggianti, B.~M., et al.\ 2009, \apjl, 697, L137
\bibitem[Postman et al.(2005)]{Postman2005}
  Postman, M., et al.\ 2005, \apj, 623, 721
\bibitem[Pratt(1993)]{Pratt1993}
  Pratt L. Y., et al.\ 1993, in Advances in Neural Information Proc. Systems 5, 204
\bibitem[Rood and Sastry(1971)]{Rood1971}
  Rood, H.~J., \& Sastry, G.~N.\ 1971, \pasp, 83, 313
\bibitem[Russakovsky et al.(2014)]{Russakovsky2014}
  Russakovsky, O., et al.\ 2014, arXiv e-prints, arXiv:1409.0575
\bibitem[Rykoff et al.(2014)]{Rykoff2014}
  Rykoff, E.~S., et al.\ 2014, \apj, 785, 104
\bibitem[Schawinski et al.(2014)]{Schawinski2014}
  Schawinski, K., et al.\ 2014, \mnras, 440, 889
\bibitem[Scoville et al.(2007)]{Scoville2007}
  Scoville, N., et al.\ 2007, \apjs, 172, 1
\bibitem[Shimakawa et al.(2021)]{Shimakawa2021}
  Shimakawa, R., et al.\ 2021, \mnras,
\bibitem[Simonyan and Zisserman(2014)]{Simonyan2014}
  Simonyan, K., \& Zisserman, A.\ 2014, arXiv e-prints, arXiv:1409.1556
\bibitem[Skibba et al.(2009)]{Skibba2009}
  Skibba, R.~A., et al.\ 2009, \mnras, 399, 966
\bibitem[Smith et al.(2005)]{Smith2005}
  Smith, G.~P., et al.\ 2005, \apj, 620, 78
\bibitem[Somerville and Dav{\'e}(2015)]{Somerville2015}
  Somerville, R.~S., \& Dav{\'e}, R.\ 2015, \araa, 53, 51
\bibitem[Spergel et al.(2015)]{Spergel2015}
  Spergel, D., et al.\ 2015, arXiv e-prints, arXiv:1503.03757
\bibitem[Szalay et al.(2001)]{Szalay2001}
  Szalay, A., et al.\ 2001, arXiv e-prints, cs/0111015
\bibitem[Tadaki et al.(2020)]{Tadaki2020}
  Tadaki, K.-. ichi ., et al.\ 2020, \mnras, 496, 4276
\bibitem[Takada et al.(2014)]{Takada2014}
  Takada, M., et al.\ 2014, \pasj, 66, R1
\bibitem[Tamura et al.(2016)]{Tamura2016}
  Tamura, N., et al.\ 2016, \procspie, 9908, 99081M
\bibitem[Tan et al.(2018)]{Tan2018}
  Tan, C., et al.\ 2018, arXiv e-prints, arXiv:1808.01974
\bibitem[Tanaka et al.(2005)]{Tanaka2005}
  Tanaka, M., et al.\ 2005, \mnras, 362, 268
\bibitem[Tanaka(2015)]{Tanaka2015}
  Tanaka, M.\ 2015, \apj, 801, 20
\bibitem[Tanaka et al.(2018)]{Tanaka2018}
  Tanaka, M., et al.\ 2018, \pasj, 70, S9
\bibitem[Tang et al.(2019)]{Tang2019}
  Tang, H., Scaife, A.~M.~M., \& Leahy, J.~P.\ 2019, \mnras, 488, 3358
\bibitem[Tasca et al.(2009)]{Tasca2009}
  Tasca, L.~A.~M., et al.\ 2009, \aap, 503, 379
\bibitem[Taylor(2005)]{Taylor2005}
  Taylor, M.~B.\ 2005, Astronomical Data Analysis Software and Systems XIV, 347, 29
\bibitem[Terlevich et al.(2001)]{Terlevich2001}
  Terlevich, A.~I., Caldwell, N., \& Bower, R.~G.\ 2001, \mnras, 326, 1547
\bibitem[van den Bergh(1976)]{Bergh1976}
  van den Bergh, S.\ 1976, \apj, 206, 883
\bibitem[van der Wel(2008)]{Wel2008}
  van der Wel, A.\ 2008, \apjl, 675, L13
\bibitem[Veale et al.(2017)]{Veale2017}
  Veale, M., et al.\ 2017, \mnras, 471, 1428
\bibitem[Visvanathan and Sandage(1977)]{Visvanathan1977}
  Visvanathan, N., \& Sandage, A.\ 1977, \apj, 216, 214
\bibitem[Vulcani et al.(2011)]{Vulcani2011}
  Vulcani, B., et al.\ 2011, \mnras, 412, 246
\bibitem[Wetzel et al.(2012)]{Wetzel2012}
  Wetzel, A.~R., Tinker, J.~L., \& Conroy, C.\ 2012, \mnras, 424, 232
\bibitem[Wetzel et al.(2013)]{Wetzel2013}
  Wetzel, A.~R., et al.\ 2013, \mnras, 432, 336
\bibitem[White et al.(2005)]{White2005}
  White, S.~D.~M., et al.\ 2005, \aap, 444, 365
\bibitem[White and Frenk(1991)]{White1991}
  White, S.~D.~M., \& Frenk, C.~S.\ 1991, \apj, 379, 52
\bibitem[Willett et al.(2017)]{Willett2017}
  Willett, K.~W., et al.\ 2017, \mnras, 464, 4176
\bibitem[Wolf et al.(2007)]{Wolf2007}
  Wolf, C., et al.\ 2007, \mnras, 376, L1
\bibitem[Wolf et al.(2009)]{Wolf2009}
  Wolf, C., et al.\ 2009, \mnras, 393, 1302
\bibitem[York et al.(2000)]{York2000}
  York, D.~G., et al.\ 2000, \aj, 120, 1579
\bibitem[Zhu et al.(2019)]{Zhu2019}
  Zhu, X.-P., et al.\ 2019, \apss, 364, 55
\end{thebibliography}
\end{document}